\documentclass[a4paper,11pt]{article}
\pdfoutput=1 

\usepackage{jheppub} 

\usepackage[T1]{fontenc} 

\font\mybb=msbm10 at 12pt

\font\myeu=eufm10 at 12pt
\def\bb#1{\hbox{\mybb#1}}

\def\frak#1{\hbox{\myeu#1}}
\def\ZZ {\bb{Z}}

\def\PP {\bb{P}}

\def\g {\frak{g}}

\def\CC{{\bf C}}

\def\Fn{{\bf F}_n}

\newcommand\beqa{\begin{eqnarray}}
\newcommand\eeqa{\end{eqnarray}}
\newcommand\n{\nonumber\\}

\begin{document}

{~}

\title{Anomaly-free Multiple Singularity Enhancement in F-theory}

\author[a,b]{Shun'ya Mizoguchi,\note{E-mail:mizoguch@post.kek.jp}}
\author[c]{Taro Tani\note{E-mail:tani@kurume-nct.ac.jp}}

\affiliation[a]{Theory Center, Institute of Particle and Nuclear Studies, KEK\\
Tsukuba, Ibaraki, 305-0801, Japan}
\affiliation[b]{Graduate University for Advanced Studies (Sokendai)}
\affiliation[c]{Kurume National College of Technology, \\Kurume, Fukuoka, 830-8555, Japan}

\emailAdd{mizoguch@post.kek.jp}
\emailAdd{tani@kurume-nct.ac.jp}

\abstract{
We study global Calabi-Yau realizations of multiple singularity enhancement 
 relevant for family-unification model building  in F-theory.
We examine the conditions under which the generation of extra chiral 
matter at multiple singularities on 7-branes in six-dimensional 
F-theory can be consistent with anomaly cancellation. 
It is shown that the generation of  
extra matter is consistent only if it is 
accompanied by simultaneous degenerations of loci of the leading polynomial 
of the discriminant  so that the total number of chiral matter does not change.
We also show that the number of singlets expected to arise 
matches the decrease of the complex structure moduli for the restricted 
geometry.
}

\preprint{KEK-TH-1857}
\date{August 29, 2015}
\maketitle

\section{Introduction}
In \cite{FFamilyUnification}, it was pointed out that 
multiple singularities on 7-branes in F-theory 
\cite{EvidenceforFtheory} may  
serve as the basis for a realization of the coset sigma model 
spectrum relevant for ``family unification" 
\cite{BPY,BLPY,OngPRD27,Ong,KY,BKMU,IY,Buchmuller:1985rc,IKK,YanagidaYasui}. 
The key observation was 
that, in six dimensions, the representation of chiral matter 
localized at an enhanced split-type singularity \cite{MorrisonVafa,BIKMSV} 
is labeled by some 
homogeneous K\"ahler manifold, the reason for which was 
explained \cite{KatzVafa,Tani} by investigating the string junctions 
\cite{DHIZ,GZ,stringjunction1,stringjunction2,stringjunction3,stringjunction4,
FYY,stringjunction6,stringjunction7,GHS1,GHS2} near the singularity.
Applying the same argument to the singularities where multiple matter 
branes simultaneously intersect the gauge 7-branes, it was argued 
that the chiral matter hypermultiplet spectrum 
at such a multiple singularity consists of those 
that form a homogeneous K\"ahler manifold 
with more than one $U(1)$ factors in the denominator of the coset.

In this paper, we examine whether this chiral matter spectrum at 
such a multiple singularity 
is 
consistent with the absence of anomalies of the theory\footnote{Singularity 
enhancement with rank more than one was considered in 
the extensive study \cite{MorrisonTaylor} 
on singularities and matter representations. 
}.
In six dimensions the condition for anomaly cancellation 
imposes a severe restriction on the chiral matter 
spectrum \cite{GSWest,KumarMorrisonTaylor1,KumarMorrisonTaylor2}.\footnote{The anomaly analysis 
has also been useful for the study of six-dimensional conformal field 
theories (See e.g. \cite{Harvey:1998bx,Intriligator:2000eq,Yi:2001bz,Ohmori:2014pca,Ohmori:2014kda}).
}
At first sight, it seems that the matter spectrum corresponding to a 
coset with multiple $U(1)$ factors 
conflicts with anomaly cancellation since it needs to be 
accompanied by generation of extra chiral hypermultiplets. 
We will, however, show that such a coset spectrum is indeed 
possible without ruining the balance of anomalies. Rather, 
in some cases when the complex structure moduli take certain values,
the absence of anomalies {\em requires} that there must occur such 
generation of extra matter at the multiple singularity.
Although we work in the F-theory compactifications on elliptic Calabi-Yau 
threefolds over a Hirzebruch surface $\Fn$ \cite{MorrisonVafa,BIKMSV}, 
the best understood example 
of an F-theory compactification, the mechanism we find is local and 
will apply to other compactifications on elliptic Calabi-Yau manifolds.

In the next section we recall what representations of hypermultiplets are 
expected to arise at a multiple singularity. In section 3 we review the 
anomaly cancellation mechanisms for ${\cal N}=1$, $D=6$ supersymmetric 
theories.  We will also see there that in the case of six-dimensional F-theory 
on an elliptic CY3 over $\Fn$, which is known to be dual to $E_8\times E_8$ 
heterotic string on $K3$,  no net increase of chiral matter is allowed 
either by the ordinary heterotic Green-Schwarz mechanism or by 
the generalized Green-Schwarz mechanism first applied by Sadov.
In section 4, we give examples of 7-brane configurations which include 
some multiple singularities {\em but} the number of hypermultiplets in 
each representation {\em does not change}  compared with the generic 
7-brane configurations in the nearby moduli space, and hence the theory 
remains anomaly-free. 
As we will see, such a transition is possible if and only if 
it is accompanied by simultaneous degenerations of loci of the leading polynomial 
of the discriminant so that the necessary extra degrees of freedom at the singularity
may be supplemented by the appropriate number of ``extra-zero" loci 
joining there simultaneously. We will also show that the decrease of the dimensions of 
the moduli space for the special class of configurations matches the number 
of new singlets appearing at the multiple singularity, which is consistent 
with the anomaly cancellation.

\section{Multiple singularity enhancement in F-theory in six dimensions}
\subsection{F-theory on an elliptic CY3 over $\Fn$}
Let us recall the basic setting of the F-theory compactification on an elliptically 
fibered Calabi-Yau over a Hirzebruch surface $\Fn$  \cite{MorrisonVafa,BIKMSV}.
The three-fold is defined by the Weierstrass equation:
\beqa
y^2&=&x^3 + f(z,z') x + g(z,z'),\label{Weierstrass}\\
f(z,z')&=&\sum_{i=0}^8 z^i f_{8+(4-i)n}(z'),\rule{0pt}{28pt}
\label{genericf}\\
g(z,z')&=&\sum_{i=0}^{12} z^i g_{12+(6-i)n}(z').\label{genericg}
\eeqa
A Hirzebruch surface $\Fn$ is a $\PP^1$ bundle over $\PP^1$. 
$z$ and $z'$ is the coordinates of the fiber and the base, respectively.
The coefficients $f_{8+(4-i)n}(z')$ $(i=0,\ldots,6)$ and 
$g_{12+(6-i)n}(z')$ $(i=0,\ldots,12)$ are polynomials 
of $z'$ of degrees specified by the subscripts. Both $x$ and $y$ are 
complex, so the equation (\ref{Weierstrass}) determines some torus 
at each $(z,z')$. More precisely, $x$,$y$,$f$ and $g$ are sections of 
${\cal L}^2,{\cal L}^3,{\cal L}^4$ and ${\cal L}^6$, where ${\cal L}$ is the 
anti-canonical line bundle of the base $\Fn$.  The total space 
is then an elliptic Calabi-Yau threefold, which is also a $K3$ fiberation 
over the $\PP^1$ parameterized by $z'$.

In order to illustrate what kind of singularity we are interested in, let us first 
consider a concrete example. 
Suppose that 
the coefficient polynomials 
of the lower order terms in the expansions of 
$f(z,z')$ (\ref{genericf}) and $g(z,z')$ (\ref{genericg}) take  
the particular forms:
\beqa
f_{4n+8}&=&-3 h_{n+2}^4,
\n
f_{3n+8}&=&12  h_{n+2}^2 H_{n+4},
\n
f_{2n+8}&=& 12\left(h_{n+2} q_{n+6}-
   H_{n+4}^2\right),\n
g_{6n+12}&=&
2 h_{n+2}^6,\n
g_{5n+12}&=&-12  h_{n+2}^4 H_{n+4},
\n
g_{4n+12}&=&12  h_{n+2}^2(2 
   H_{n+4}^2- h_{n+2} q_{n+6}),
\n
g_{3n+12}
   &=&-f_{n+8} h_{n+2}^2+24 h_{n+2}
   H_{n+4} q_{n+6}-16 H_{n+4}^3,
\n
g_{2n+12}&=&-f_8 h_{n+2}^2+2 f_{n+8}
   H_{n+4}+12 q_{n+6}^2
\label{genericSU(5)fandg}
\eeqa 
for some polynomials $h_{n+2}$,$H_{n+4}$ and $q_{n+6}$; 
they are so arranged  
that the discriminant starts with the $z^5$ term to produce a $I_5=SU(5)$ 
Kodaira singularity \cite{Kodaira} 
along the line $z=0$. 
For later convenience we present an explicit form of the lower order 
expansions of this curve in appendix A.

The independent polynomials preserving this particular singularity 
structure are
\beqa
\mbox{$h_{n+2}$, $H_{n+4}$, $q_{n+6}$, $f_{n+8}$ and $g_{n+12}$.} 
\label{SU(5)polynomials}
\eeqa
The total degrees of freedom is thus 
\beqa
(n+3)+(n+5)+(n+7)+(n+9)+(n+13)-1&=&5n+36,
\label{sumofdegrees}
\eeqa
which matches the number of $SU(5)$ singlets computed by using the index 
theorem on the heterotic side\footnote{
Note that the ``middle" coefficients $f_8, g_{12}$ and the higher ones 
$f_{8-n},\ldots$; $g_{12-n},\ldots$ are not counted here as the complex structure 
moduli which are to be compared with the singlets arising form ``this" $E_8$ 
factor. This is because the middle ones $f_8, g_{12}$ correspond to the geometric 
moduli of the elliptic $K3$ while the higher ones are taken into account 
in the similar analysis for the singularity at $z=\infty$ corresponding to the 
other (partially broken) $E_8$ gauge factor.}.

Since the leading order term of the discriminant $\Delta$ is
\beqa
\Delta&=&108 z^5 h_{n+2}^4 P_{3n+16}+\cdots,
\n
P_{3n+16}&\equiv&
-2 f_8 h_{n+2}^2 H_{n+4}-2 f_{n+8} h_{n+2}
   q_{n+6}+f_{8-n} h_{n+2}^4+g_{n+12} h_{n+2}^2-24 H_{n+4} q_{n+6}^2,
   \label{genericSU(5)Delta}
\eeqa
the singularity gets enhanced to a higher one 
at the $n+2$ zero loci of $h_{n+2}$ and 
the $3n+16$ loci of $P_{3n+16}$\footnote{Although it contains $f_{8}$ and $f_{8-n}$,
they only affect the positions of the loci and do not affect the total number of 
the loci.}.

At the zero loci of a {\em generic} $P_{3n+16}$ (so that $h_{n+2}\neq 0$, in particular),
the order of $\Delta$ becomes $\geq 6$ while ${\rm ord}f$ and ${\rm ord}g$ remain 
zero. If ${\rm ord}\Delta=6$, the singularity is enhanced to $I_6=SU(6)$ and 
a chiral matter in ${\bf 5}$ appears at each zero locus of $P_{3n+16}$. 
On the other hand, 
at the $n+2$ loci of $h_{n+2}$, the first few terms of $f$ and $g$ simultaneously 
vanish so that $f$ starts with $z^2$ and $g$ does with $z^3$, 
 as long as $H_{n+4}$ does not vanish there. Also, the order of the discriminant 
 becomes 7.
 This is the $I^*_5=SO(10)$
singularity, and the chiral matter is ${\bf 10}$ at each zero of $h_{n+2}$ for 
generic $P_{3n+16}$.  In all, the matter spectrum for the generic $SU(5)$ curve is
\beqa
(n+2){\bf 10},~~(3n+16){\bf 5},~~(5n+36){\bf 1}.
\label{SU(5)curvespectrum}
\eeqa

Originally \cite{BIKMSV} what kind of charged matter should appear at these enhanced 
``extra zeroes'' was determined by referring to the massless spectrum of the 
dual heterotic model \cite{MorrisonVafa}, that is, the $K3$ compactification of the 
$E_8\times E_8$
heterotic string with instanton numbers $(12-n, 12+n)$.
%
%
%
The relationship between the 
extra zeroes of the discriminant and the massless charged matter was 
first explained 
by Katz and Vafa \cite{KatzVafa} by mapping the problem to that of 
deformations of the singularities of $K3$.
Later it was proposed by one of the present authors \cite{Tani} 
how the chiral matter spectrum is understood by 
investigating string junctions near the enhanced singularity.\\

%

\noindent
{\it Spectral cover, matter localization and the Mordell-Weil group}

One of the remarkable features of heterotic/F-theory duality is that 
 a brane-like object naturally comes into play in 
heterotic theory through the construction of a vector bundle 
over the elliptic Calabi-Yau manifold \cite{FMW}.
Basically, the statement of heterotic/F-theory duality is made 
in a certain limit in the moduli space on both sides: 
F-theory is compactified on a $K3$-fibered Calabi-Yau  
where the $K3$ goes to a stable degeneration limit into two,  
themselves elliptically fibered, $dP_9$'s 
intersecting along a two torus $E$, and heterotic string theory is 
on an elliptically fibered Calabi-Yau whose fiber torus has a 
large volume and the same complex structure as $E$. 
The moduli space of the vector bundle over each torus is known 
as Looijgenha's weighted projective space; for an $SU(5)$ 
gauge group this is an ordinary projective space.
The {\em spectral cover} is a polynomial equation of $x$ and $y$, 
the variables in the Weierstrass equation describing a 
heterotic torus fiber. The defining polynomial has five (for $SU(5)$) 
zero loci (which add up to zero) on the torus, each of which 
specifies a Wilson line of a Cartan generator and coordinatizes 
Looijgenha's projective space. In the $SU(5)$ case, the polynomial 
is explicitly \cite{FMW}:
\beqa
w&=&a_0+a_2 x+a_3 y+a_4 x^2+a_5 x^2 y
\label{w}
\eeqa
for some coefficients $a_0,\ldots,a_5$.

On the other hand, we consider a pencil \cite{DonagiWijnholt}
\beqa
&&(y^2+x^3+\alpha_1 x y z +\alpha_2 x^2 z^2 
+\alpha_3 y v^3 +\alpha_4 x v^4 +\alpha_6 v^6)+p(v,x,y)u=0,
\label{pencil}\\
&&p(v,x,y)~=~a_0 v^5+a_2 x v^3+a_3 y v^2+a_4 x^2 v+a_5 x^2 y
\label{p}
\eeqa
in ${\bf WP}^3_{(1,1,2,3)}$ with the equivalence relation 
$(u,v,x,y)\sim (\lambda u, \lambda v, \lambda^2 x, \lambda^3 y)$, 
$\lambda\in \CC$. Obviously, $p(v,x,y)$ (\ref{p}) is the homogenization 
of $w$ (\ref{w}). After blowing up $u=v=0$ the pencil (\ref{pencil}) 
becomes $dP_9$, which we regard as one of two $dP_9$'s appearing 
in the stable degeneration limit on the F-theory side. Indeed, we
can show that if we set
\beqa
a_5
&=& 2 \sqrt{3} u^{-1} h_{n+2},\n
a_4
&=& u^{-1}(\sqrt{3} \alpha_1 h_{n+2}+6
   H_{n+4}),\n
a_3
&=&-\sqrt{3} u^{-1}\left({\textstyle \frac16} (\alpha_1^2 -4 \alpha_2)h_{n+2}
+4 q_{n+6}\right),\n
a_2
&=&
u^{-1}\left(
\sqrt{3}(-{\textstyle \frac1{12}} \alpha_1^3 +{\textstyle \frac13}
    \alpha_1 \alpha_2 + \alpha_3 )h_{n+2}
+(- \alpha_1^2 +4 \alpha_2)
   H_{n+4}
-2 \sqrt{3} \alpha_1 q_{n+6}+ f_{n+8}\right),\n
a_0
&=&u^{-1}\left(
{\textstyle\frac{\sqrt{3}}{12}} \alpha_3
\left(4 
   \alpha_2 -\alpha_1^2
   \right)
h_{n+2} 
+ (2\alpha_4 - \alpha_1\alpha_3 ) H_{n+4}
-2 \sqrt{3} \alpha_3 q_{n+6}
+
{\textstyle\frac1{12}}(4 \alpha_2-\alpha_1^2)
   f_{n+8}
-g_{n+12}
\right),\n
\eeqa
the pencil (\ref{pencil}) precisely reproduces the lower terms 
(up to the ``middle'' ones) of the $SU(5)$ Weierstrass equation 
(\ref{Weierstrass})(\ref{genericf})(\ref{genericg}) 
with (\ref{genericSU(5)fandg}).
Therefore, the polynomials (\ref{SU(5)polynomials}) of \cite{BIKMSV} 
correspond to
\beqa
h_{n+2}\sim a_5,~~ H_{n+4}\sim a_4,~~ q_{n+6}\sim a_3, ~~
f_{n+8}\sim a_2, ~~g_{n+12}\sim a_0.
\label{h's_and_a's}
\eeqa

Furthermore, it was also shown by using the Leray spectral sequence 
\cite{Curio,DiaconescuIonesei} 
that the matter is localized where some of $a_j$'s vanish and some 
of the zero loci of $w$ (or $p(v,x,y)$) go to infinity. We note that
this may be intuitively understood as a consequence of the structure 
theorem of the Mordell-Weil group \cite{Aspinwall,AspinwallMorrison}. 
Indeed, the equation $p(v,x,y)=0$ 
defines sections of $dP_9$, and since the structure theorem 
\cite{Shioda,OguisoShioda} states the singularities and the sections are 
orthogonal complement of each other in $E_8$, 
the less sections we have, the more singularities we get instead.

The Mordell-Weil lattice was studied in detail 
in terms of string junctions in \cite{FYY} using the isomorphism between 
the string junction algebra and the Picard lattice of a rational elliptic surface.
For a recent F-theory phenomenological aspect of the Mordell-Weil group 
see \cite{Lawrie:2015hia,Krippendorf:2015kta}.

\begin{figure}[b]%
\mbox{\hskip -15ex
\includegraphics[height=0.40\textheight]{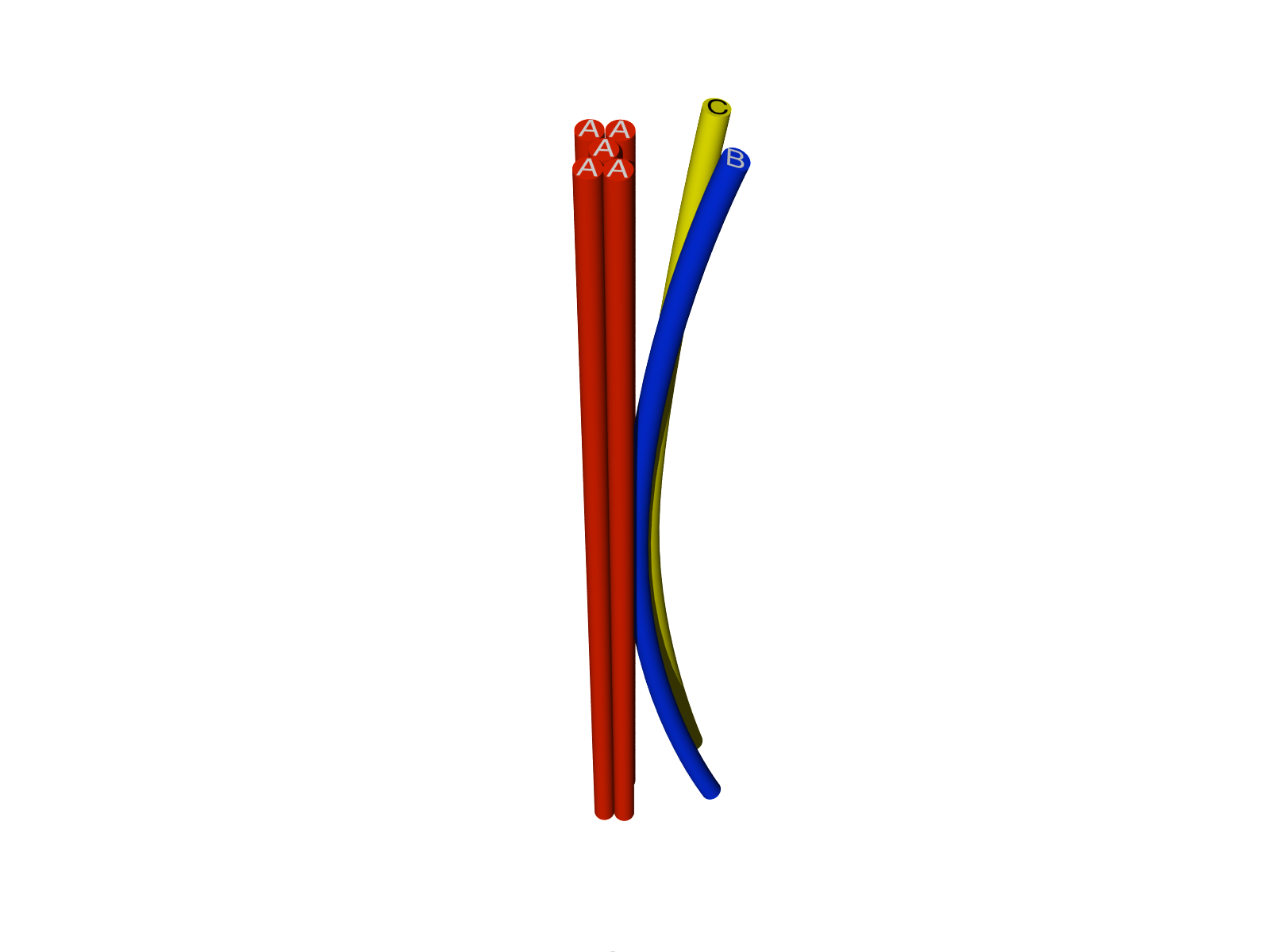}\hskip -35ex
\includegraphics[height=0.40\textheight]{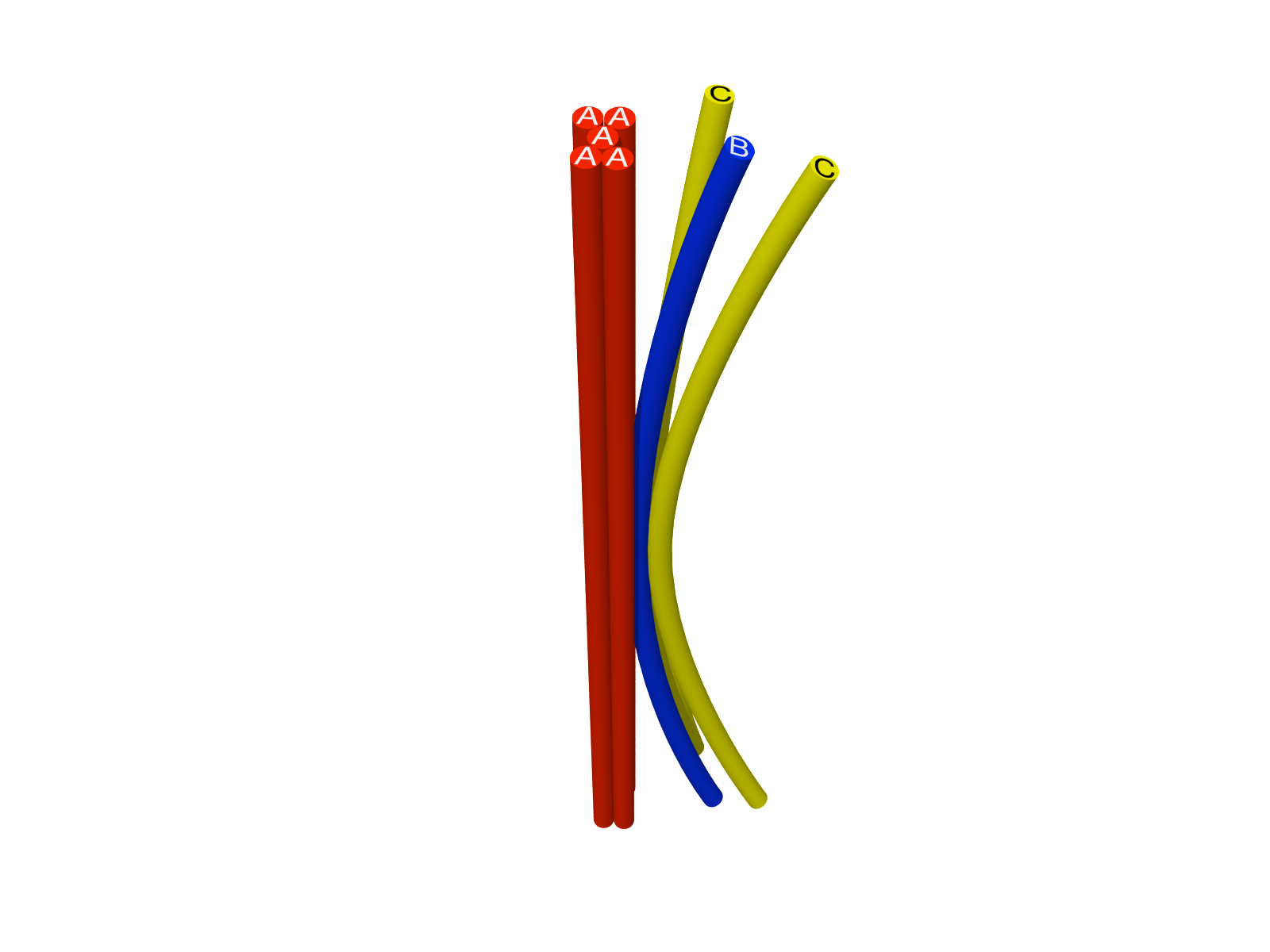}
}
\\
\caption{\label{} Left: $D_5$ singularity. Right: $E_6$ singularity. }
\end{figure}

\subsection{Multiple singularity enhancement from $SU(5)$ to $E_6$}
We will now consider what happens if $h_{n+2}$ and $H_{n+4}$  
simultaneously vanish. 
In this case, the $z^2$ term of $f$ and the $z^3$ term of $g$ 
vanish at these points, and the order of the discriminant rises up to 8. 
This means that the singularity gets enhanced from $I_5=SU(5)$ to 
$IV^*=E_6$ there. 

Note that if $h_{n+2}=0$, that $H_{n+4}$ vanishes means that $P_{3n+16}$ 
also does. Thus this higher singularity can be viewed as a consequence 
of a collision of an $I^*_1=SO(10)$ singularity, occurring at a zero of $h_{n+2}$,
and an $I_6=SU(6)$ singularity, which corresponds to a zero of  $P_{3n+16}$. 

 In the standard 7-brane 
 representation of the Kodaira singularity, the $SU(5)$ singularity is made of a 
 collection of five {\bf A}-branes, while the $SO(10)$ singularity is 
 represented by ${\bf A}^5{\bf B}{\bf C}$.
 Thus the zero loci of the polynomial $h_{n+2}$ are the places where
 a {\bf B}- and a {\bf C}-branes intersect the five {\bf A}-branes 
 lying on top of each other (FIG.1, left).
On the other hand, if $H_{n+4}$ happens to vanish at the same point, 
then the singularity becomes $E_6$ which is 
represented by ${\bf A}^5{\bf B}{\bf C}{\bf C}$.
Therefore, this multiple singularity occurs when an extra 
{\bf C}-brane simultaneously meets the five {\bf A}-branes together in addition 
to  the {\bf B}- and {\bf C}-branes (FIG.1, right).

However, suppose that we slightly move from this special point in the 
complex structure moduli space to another where $h_{n+2}$ and $H_{n+4}$ 
do {\em not} simultaneously vanish but the roots of 
$h_{n+2}=0$ and $P_{3n+16}=0$ are still close (Note that if $h_{n+2}$ is not 
zero, $H_{n+4}=0$ does not mean $P_{3n+16}=0$.) This will correspond to 
the split of the multiple $E_6$ singularity into an $SO(10)$ singularity and 
an $SU(6)$ singularity. 
While it is OK for the pair of  {\bf B}- and {\bf C}-branes to form the $D_5$ 
singularity,  how can the remaining {\bf C}-brane yield the $A_5$ 
singularity with the five  {\bf A}-branes?

\begin{figure}[b]%
\mbox{\hskip 0ex
\includegraphics[height=0.45\textheight]{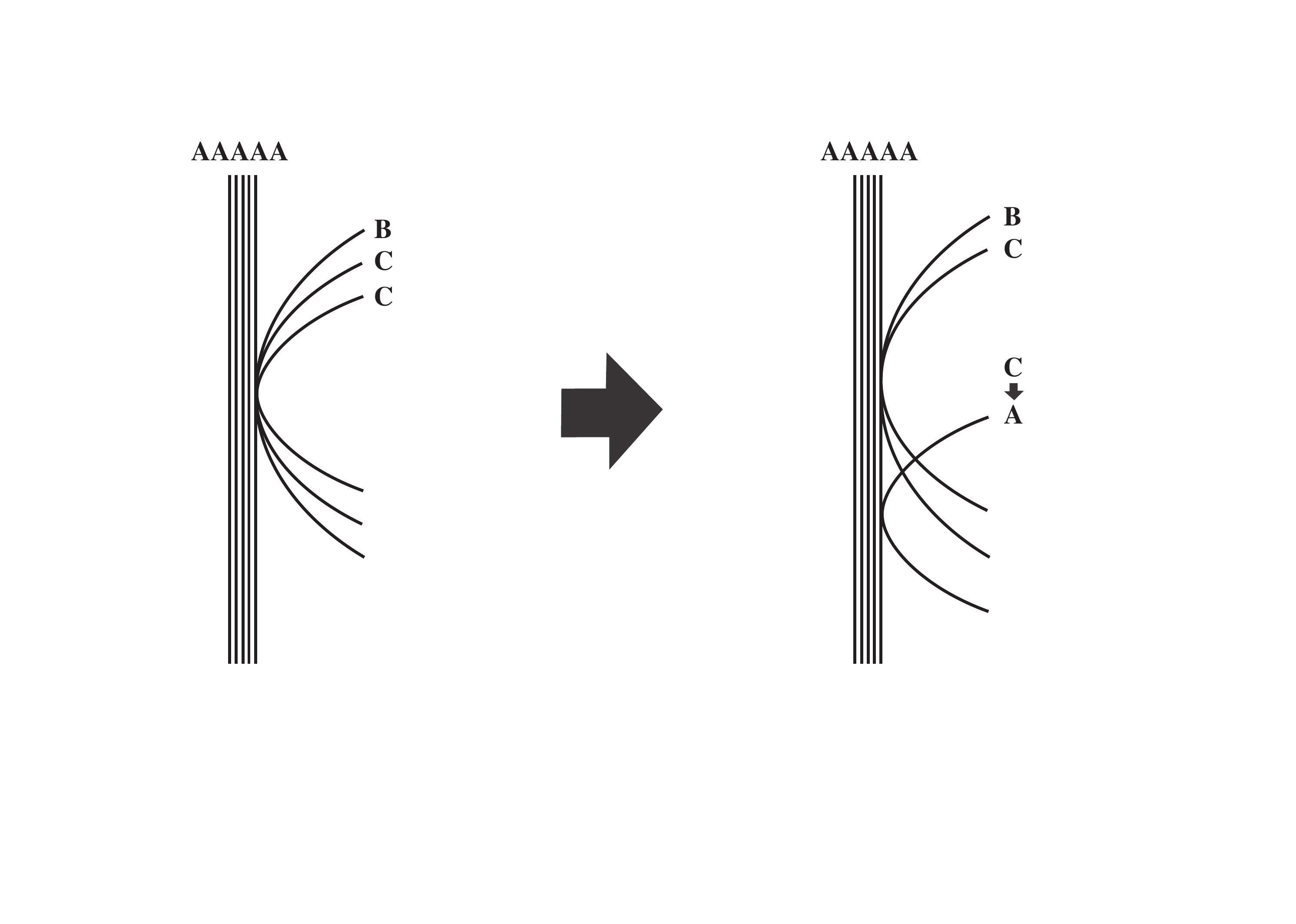}
}
\\
\caption{\label{} An $E_6$ singularity is split into a $D_5$ and an $A_5$ singularities. }
\end{figure}

This apparent contradiction can be explained as follows:
We should first note that {\em any} isolated discriminant locus has 
monodromy 
\beqa
T&=&\left(
\begin{array}{cc}
1&~~1\\0&~~1
\end{array}
\right)
\eeqa
  and hence, {\rm locally}, is identified as a location of an 
{\bf A}-brane. However, when 
this discriminant locus merges with 
a $D_5$ singularity to form an $E_6$ singularity, 
a pair of zero loci of $g$ and $f$ also joins with the discriminant locus 
since the orders of $g$ and $f$ are respectively enhanced by one.

To understand why the {\bf C}-brane can produce the $A_5$ singularity, 
we must know the monodromies around the zero loci of $f$ and $g$. 
The zero locus of $f$ is mapped, by the inverse $J$-function via 
the relation 
\beqa
J(\tau)&=&\frac{4f^3}{4f^3 + 27 g^2},
 \eeqa 
 to (taking the starting point in $\mbox{Im} J>0$ in
 the standard fundamental region of the modular group) 
 $\tau=e^{\frac{2\pi i}3}$, near which $ J(\tau)$ behaves like
 \beqa
 J(\tau)&=&(\tau-e^{\frac{2\pi i}3})^3(1+ O(\tau-e^{\frac{2\pi i}3})).
 \eeqa
That is, if $J$ changes its value along a small closed path 
encircling $0$ {\em three times}, $\tau$ goes 
around $e^{\frac{2\pi i}3}$ precisely once, back to the original fundamental 
region; this can be verified by tracing the value of the $J$ function 
\cite{Tani}: 
Since $J\simeq\mbox{const.}f^3$ near $f=0$,
if one goes around the zero locus of $f$ once counter-clockwise on the $z$ plane, 
the value of $J$ goes around zero three times 
counter-clockwise. 
Therefore, the monodromy around the locus of $f$ is
\beqa
(ST^{-1})^3&=&-1\n
&\simeq&1~~~\mbox{in $PSL(2,\ZZ)$},
\eeqa
and hence is identity as a modular transformation.
Here 
\beqa
S&=&
\left(
\begin{array}{cc}
0&~-1\\1&~~0
\end{array}
\right).
\eeqa

\begin{figure}[b]%
\mbox{\hskip 0ex
\includegraphics[height=0.50\textheight]{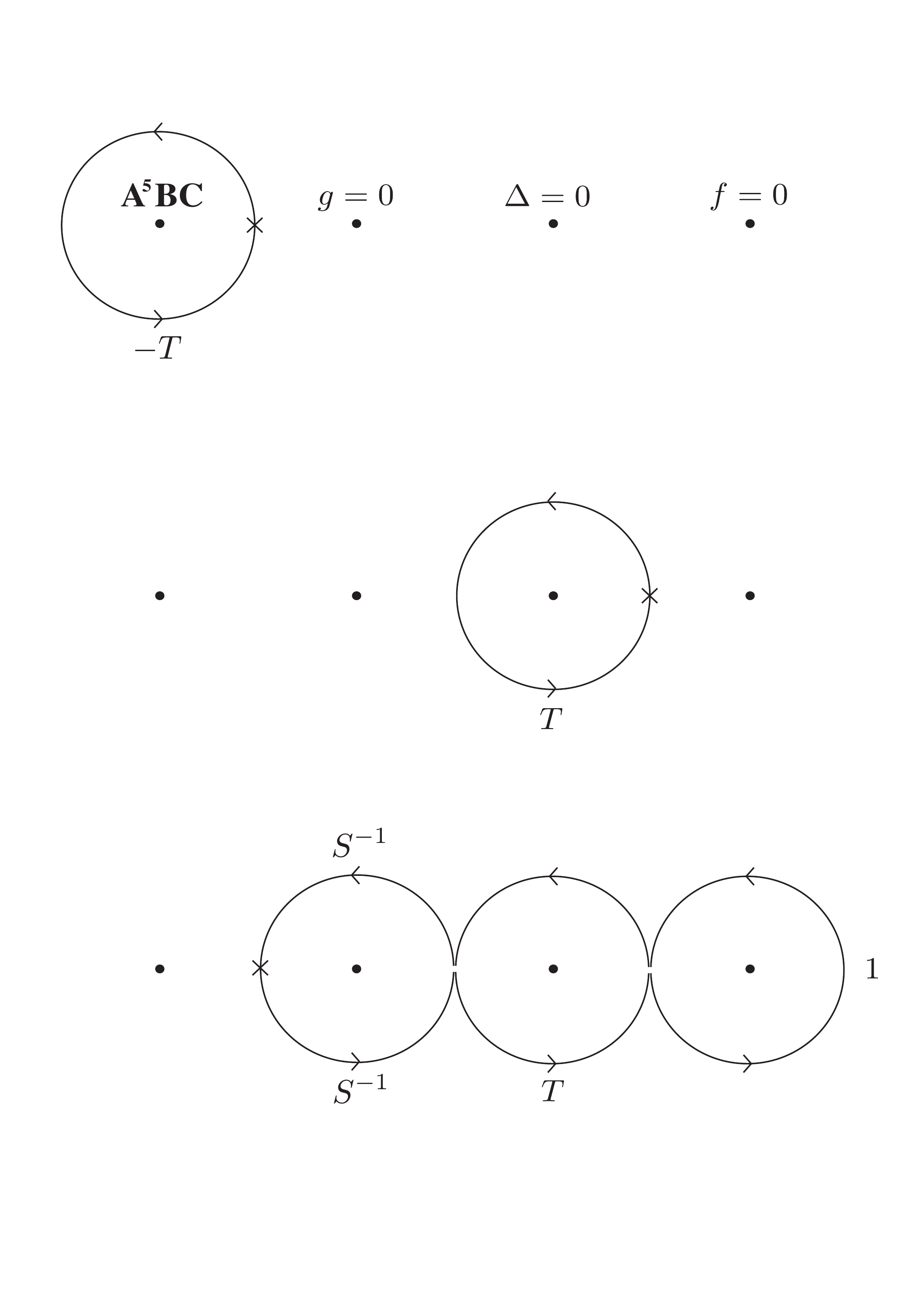}}
\\
\caption{\label{monodromies}  }
\end{figure}


Likewise, the zero locus of $g$ is mapped to $\tau = i$, and 
the expansion of $J(\tau)$ is then
\beqa
 J(\tau)&=&1+(\tau-i)^2(1+ O(\tau-i)).
\eeqa
So if the value of $J$ goes around $1$ {\em twice}, 
$\tau$ then does around $i$ once, again back to the original fundamental 
region.
Since $J-1\simeq\mbox{const.}g^2$ near $g=0$, circling around the zero locus 
of $g$ once on the $z$ plane means that the corresponding $\tau$ 
circles around $i$ once. Thus the monodromy around the locus of $g$ reads
\beqa
(S^{-1})^2&=&-1\n
&\simeq&1~~~\mbox{in $PSL(2,\ZZ)$},
\eeqa
and again is identity.

\begin{table}[bh]

\caption{Singularities and string junctions for the unbroken $SU(5)$ case. \label{singularityandstringjunctions}}
\centering
\begin{tabular}{|c|c|c|c|c|}
\hline
&singularity& 7-brane &string junction &$SU(5)$ representation\\
\hline
generic $z'$
&\hskip -10ex\raisebox{-10ex}{\includegraphics[height=0.15\textheight]{A4only_image.png}}
\hskip -10ex$I_5$
&${\bf A}^5$ &$\begin{array}{c}\pm(a_i-a_j)~(i<j)\\
\mbox{+ Cartan}\end{array}$& 
$\begin{array}{c}{\bf 24}\\\mbox{(gauge symmetry)}\end{array}$
\\
\hline
\hline
locus of $h_{n+2}$
&\hskip -10ex\raisebox{-10ex}{\includegraphics[height=0.15\textheight]{D5toA4_image.png}}
\hskip -10ex$I^*_1$ 
&${\bf A}^5{\bf B}{\bf C}$  &$a_i+a_j-b-c ~(i<j)$& {\bf 10}
\\
\hline
locus of $P_{3n+16}$
&\hskip -8ex\raisebox{-10ex}{\includegraphics[height=0.15\textheight]{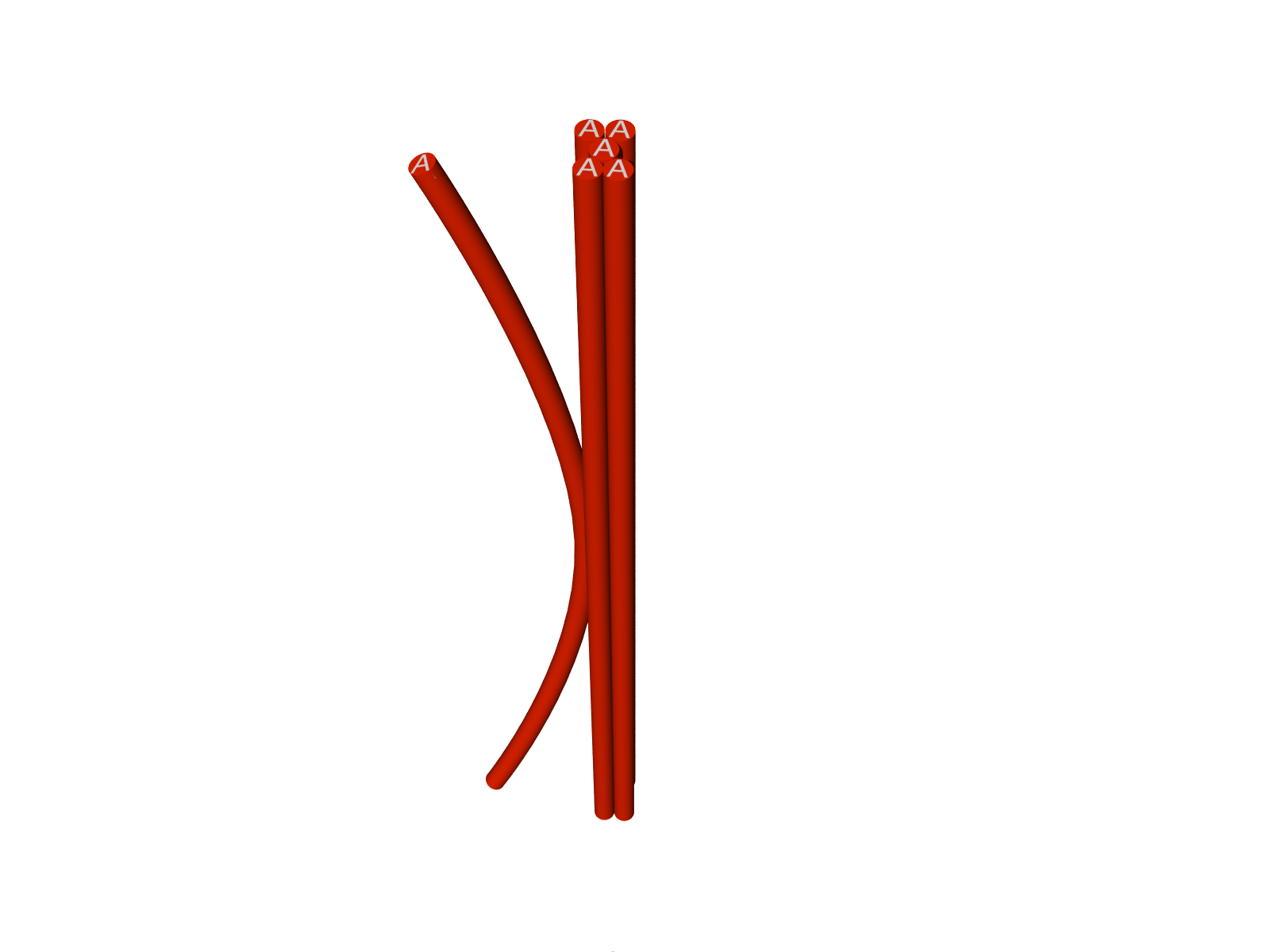}}
\hskip -10ex$I_6$ 
&${\bf A}^5{\bf A}'$  &$a_i-a'~(i=1,\ldots,5)$& {\bf 5}
\\
\hline
$\begin{array}{c}
\mbox{common locus of}\\
\mbox{$h_{n+2}$ and $P_{3n+16}$}
\end{array}$
&\hskip -10ex\raisebox{-10ex}{\includegraphics[height=0.15\textheight]{E6toA4_image.png}}
\hskip -10ex$IV^*$ 
&${\bf A}^5{\bf B}{\bf C}{\bf C}'$  &
$\begin{array}{c}
a_i+a_j-b-c ~(i<j)\\
a_i+a_j-b-c' ~(i<j)\\
\sum_{k=1}^5 a_k -a_i-2b -c -c'\\
c-c'
\end{array}$
& 
$\begin{array}{c}
{\bf 10}\\
{\bf 10}\\
{\bf 5}\\
{\bf 1}
\end{array}$
\\
\hline
\end{tabular}

\end{table}

Now let us consider the effect of these loci of $f$ and $g$ to the monodromy of 
the other coalescing 7-branes.
As we discussed above, any discriminant locus is locally an {\bf A} brane.
However, when this and the $ {\bf A}^5 {\bf B} {\bf C}$ branes come close to merge, 
it turns out that there is also a locus of $g$ situated in between them.  
So if the reference point of the monodromy is set near the $ {\bf A}^5 {\bf B} {\bf C}$ branes,
then one undergoes the $S^{-1}$ transformation when one passes by the locus of $g$.
Therefore, the monodromy of the discriminant locus is 
\beqa
S^{-1}T S^{-1}
\eeqa
(FIG. \ref{monodromies}), which is equal to 
\beqa
-T^{-1}CT &\simeq&T^{-1}CT~~~\mbox{in $PSL(2,\ZZ)$},
\eeqa
where $C$ is the monodromy matrix of {\bf C}:
\beqa
C&=&
\left(
\begin{array}{cc}
0&~1\\-1&~2
\end{array}
\right).
\eeqa
Since the monodromy matrix of $ {\bf A}^5 {\bf B} {\bf C}$ is $=-T$ which is 
invariant under $T$ conjugation, we see that this discriminant locus can 
merge with the $D_5$ singularity as a {\bf C} brane.
If, on the other hand, the {\bf B} and {\bf C} branes split off from the coalesced 
five {\bf A} branes and one ``{\bf C} brane", 
the locus of $g$ also splits off, and therefore what has been a {\bf C} brane 
in the $E_6$ singularity turns into an  {\bf A} brane, yielding the 
$A_5$ singularity together with the five {\bf A} branes.

As we saw in the previous section, if the {\bf A} brane alone meets the gauge 7-branes 
while the  {\bf B} and {\bf C} branes are apart, then this singularity corresponds to one of 
the roots of $P_{3n+16}=0$ and the charged matter ${\bf 5}$ appears. These
 BPS states are thought of as coming from string junctions connecting 
the {\bf A} brane and the gauge branes localized near the intersection point.
Likewise, if only the  {\bf B} and {\bf C} branes intersect while the {\bf A} brane is 
apart, the singularity is of $SO(10)$ and a ${\bf 10}$ will arise 
due to the string junctions connecting the the  {\bf B} and {\bf C} branes and the
gauge branes; this happens at the loci of $h_{n+2}$.
Therefore, if  $P_{3n+16}=0$ and $h_{n+2}$ are simultaneously zero, then there 
will arise both ${\bf 10}$ and ${\bf 5}$ at that point. The former comes from the 
string junctions $a_i+a_j-b-c$ $(1\leq i<j\leq 5)$ while the latter can be 
identified as the ones of the form $\sum_{k=1}^5 a_k -a_i-2b -c -c'$ 
(TABLE I)\footnote{For each matter locus there are in fact twice as many BPS junctions
corresponding to overall $\pm$ multiplications.}.

But we also notice that, at this $E_6$ point, there are not only 
these two kinds of string junctions but 
still more BPS junctions: the ones of the form 
$a_i+a_j-b-c'$ $(1\leq i<j\leq 5)$ and $c-c'$. 
They are special BPS junctions that appear only at this higher singularity and 
are not present at generic points in the moduli space.
Since they are BPS, these extra string junctions are also expected to give rise to 
chiral matter at the multiple singularity. This was the proposal of ref.\cite{FFamilyUnification}.

However, there is a puzzle here: Suppose that the theory is at some generic 
point in the moduli space of an elliptic Calabi-Yau three-fold over a Hirzebruch 
surface where it is dual to some $E_8\times E_8$ heterotic string compactification 
on $K3$. Of course, the theory is anomaly free. Then suppose that the values of 
the moduli parameters are tuned to some special ones so that the 7-branes 
develop a multiple singularity. 
If it gives rise to more chiral matter hypermultiplets than those present 
at a generic point in the moduli space, 
doesn't that conflict with anomaly cancellation?

In the next section, we will see that the requirement of anomaly cancellation severely 
limits the conditions under which this phenomenon consistently occurs. 

\section{Anomaly cancellation in six dimensions}
The relevant anomaly eight-forms are given 
by\footnote{They are $-16\pi^4$ times the ones given in \cite{GSWest}.}
\beqa
({\hat I}^{(D=6)}_{3/2})_{8}&=&\frac{49}{36}Y_4-\frac{43}{72}Y_2^2,\\
({\hat I^0}_{1/2})_{8}&=&\frac{1}{180}Y_4+\frac{1}{72}Y_2^2,\\
({\hat I^{L_i}}_{1/2})_{8}&=&\mbox{dim}L_i\left(\frac{1}{180}Y_4+\frac{1}{72}Y_2^2\right)
+\frac1{16}\left(
\frac23\mbox{tr}_{L_i}F^4-\frac16\mbox{tr}_{L_i}F^2\mbox{tr} R^2
\right),\\
({\hat I}_A)_{8}&=&\frac{7}{45}Y_4-\frac{1}{9}Y_2^2,
\eeqa
where
\beqa
Y_{2m}&\equiv&\frac12\left(
-\frac14
\right)^m
\mbox{tr}R^{2m},
\eeqa 
and $L_i$ denotes the representation of the unbroken gauge group $G$.

In general, 
the total anomaly polynomial  is given by  
\beqa
{\hat I^{\rm total}}_{8}&=&
-\left(({\hat I}^{(D=6)}_{3/2})_{8}+({\hat I}_A)_{8}\right)
+n_T\left(({\hat I}_A)_{8}+({\hat I^0}_{1/2})_{8}\right)
-\sum_\alpha 
({\hat I}^{{\rm Ad}G_\alpha}_{1/2})_{8}\n&&
+
\sum_{i} n^{
i}_H
({\hat I}^{L_i
}_{1/2})_{8}
~~+n^0_H
({\hat I^0}_{1/2})_{8},
\eeqa
where $n_T$ is the number of tensor multiplets, 
$n^{
i}_H$ is the number of massless hypermultiplets in the representation 
$L_i
$ of the unbroken gauge group 
and 
$n^0_H$ is the number of other neutral hypermultiplets not counted in $n^{
i}_H$ as 
singlets. We assume that the unbroken gauge group is 
a direct product 
$\prod_\alpha G_\alpha$.
We write 
\beqa
n_V&\equiv&\sum_\alpha \mbox{dim}G_\alpha,\n
n_H&\equiv&
\sum_{i} n^{
i}_H \mbox{dim}L_i
+n^0_H,
\eeqa
then if they satisfy the well-known relation:
\newcounter{counterHV}
\setcounter{counterHV}{\theequation}
\beqa
n_H - n_V &=& 273 - 29 n_T, \label{H-V=273-29T}
\eeqa
the $\mbox{tr}R^4$ terms cancel out and we have \cite{Sadov}
\beqa
{\hat I^{\rm total}}_{8}
%
&=&\frac{9-n_T}2 Y_2^2
-\frac1{12}Y_2\sum_\alpha(
\mbox{Tr}_\alpha F_\alpha^2
-\sum_i n_H^{\alpha i}\mbox{tr}_{L^\alpha_i}F_\alpha^2
)\n
&&
-\frac1{24}
\sum_\alpha(
\mbox{Tr}_\alpha F_\alpha^4
-\sum_i n_H^{\alpha i}\mbox{tr}_{L^\alpha_i}F_\alpha^4
)\n
&&+\frac14\sum_{\alpha<\beta}\sum_{i,j}
n_H^{\alpha i;\beta j}\mbox{tr}_{L^\alpha_i}F_\alpha^2
\mbox{tr}_{L^\beta_j}F_\beta^2,
\label{Ihattotal}
\eeqa
where, as usual, $\mbox{Tr}_\alpha$ denotes the trace taken in the 
adjoint representation of $G_\alpha$.
$n_H^{\alpha i}$ is the number of hypermultiplets in the 
representation $L^\alpha_i$  of $G_\alpha$, and  
$n_H^{\alpha i;\beta j}$ is one in $L^\alpha_i\otimes L^\beta_j$ 
of $G_\alpha\times G_\beta$.

%

It is known  \cite{MorrisonVafa,BIKMSV}  that F-theory 
compactified on an elliptic Calabi-Yau three-fold over the 
the Hirzebruch surface ${\bf F}_n$ is dual to
the $K3$ compactification of the $E_8\times E_8$
heterotic string with instanton numbers $(12+n, 12-n)$, 
so let us first recall the perturbative spectrum of the $K3$ compactifications 
of the $E_8\times E_8$ heterotic string.

Let $H^{(m)}$ $(m=1,2)$ be the gauge group of instanton in 
 $E_8^{(m)}$ with instanton number $12+(-1)^{m-1} n$, 
and $G^{(m)}$ be the maximal commutant 
in $E_8^{(m)}$. 
Let the decomposition of the adjoint of $E_8$ in the representations 
of $G^{(m)}\times H^{(m)}$ be
\beqa
{\bf 248}^{(m)}&=&\oplus_i (L^{(m)}_i\otimes C^{(m)}_i).
\eeqa
for each $m=1,2$.
Let $F_0^{(m)}$ $(m=1,2)$ be the field strength of the instanton in $H^{(m)}$, 
and define $r^{(m)}_i$ as the ratio of the traces
\beqa
\mbox{tr}_{C^{(m)}_i} F^{(m)2}_0&=&r^{(m)}_i \mbox{Tr}_{E_8} F^{(m)2}_0.
\eeqa
Then the number of hypermultiplets are given by the index theorem: 
\beqa
-n^0_H&=&-21,\n
-n^{(m)\alpha i}_H&=&\mbox{dim}C^{(m)}_i-\frac1{8\pi^2}\int_{K3}\frac12\mbox{tr}_{C^{(m)}_i} F^{(m)2}_0.
\label{indexTH}
\eeqa

Using these expressions in (\ref{Ihattotal}), one can show that \cite{GSWest}
\beqa
{\hat I^{\rm total}}_{8}&=&4\left(Y_2+\frac18(x^{(1)}+x^{(2)})\right)
\left(Y_2+\frac n{16}(x^{(1)}-x^{(2)})\right),
\label{factorizedI8}
\eeqa
where $x^{(m)}=\frac1{30}\mbox{Tr}_{E_8}F^{(m)2}$ $(m=1,2)$.
Thus the anomaly of the K3 compactification of the $E_8\times E_8$ 
heterotic string factorizes and hence can be canceled by the Green-Schwarz 
mechanism.

In F-theory, an alternative anomaly cancellation mechanism is known:
The generalized Green-Schwarz mechanism assumes that \cite{Sadov} 
the same anomaly polynomial 
(\ref{Ihattotal}) can be written in a bilinear form
\beqa
{\hat I^{\rm total}}_{8}&=&\frac 12 \Omega_{\hat i \hat j}X^{\hat i}X^{\hat j},
\label{I=OmegaXX}
\\
 X^{\hat i}&\equiv&
\frac12 a^{\hat i} \mbox{tr}R^2 + \sum_\alpha 2b_\alpha^{\hat i}\mbox{tr}F_\alpha^2
\label{X}
\eeqa
for some constants $\Omega_{\hat i \hat j}$, $a^{\hat i}$ and $b_\alpha^{\hat i}$,
where
the repeated indices ${\hat i}$, ${\hat j}$ are understood 
to be summed over 1 through the total number of $B$ fields.
The anomaly is then written as
\beqa
\int \Omega_{\hat i\hat j}\omega_2^{1\hat i}X^{\hat j}
\eeqa
with
\beqa
X^{\hat i}&=&d\omega_3^{\hat i},\\
\delta_\Lambda \omega_3^{\hat i}&=& d \omega_2^{1\hat i}(\Lambda),
\eeqa
which can be canceled by the contribution from the 
counterterm
\beqa
\int \Omega_{\hat i\hat j}B^{\hat i}X^{\hat j},
\eeqa
assuming that the anomalous transformations of the $B^{\hat i}$ fields
\beqa
\delta_\Lambda B^{\hat i}&=&
-\omega_2^{1\hat i}(\Lambda).
\eeqa

The conditions for the anomaly polynomial to be  written in the form 
(\ref{I=OmegaXX}) are summarized by the following set of equations:
\beqa
9-n_T&=&\sum_{\hat i, \hat j}
\Omega_{\hat i\hat j}a^{\hat i}a^{\hat j},
\label{anomalycondition1}
\\
\mbox{index}(\mbox{Ad}G_\alpha) 
-\sum_{\hat i}n_H^{\alpha i
}\mbox{index}(L_{i}^\alpha)&=&
6\sum_{\hat i, \hat j}
\Omega_{\hat i\hat j}a^{\hat i}b_\alpha^{\hat j}
\label{anomalycondition2}
\\
x_{\mbox{\scriptsize Ad}G_\alpha}-\sum_{\hat i}n_H^{\alpha i
}x_{L_{i}^\alpha}&=&0,
\label{anomalycondition3}
\\
y_{\mbox{\scriptsize Ad}G_\alpha}-\sum_{\hat i}n_H^{\alpha i
}y_{L_{i}^\alpha}
&=&
-3\sum_{\hat i, \hat j}
\Omega_{\hat i\hat j}b_\alpha^{\hat i}b_\alpha^{\hat j},
\label{anomalycondition4}
\\
\sum_{\hat i, \hat j}
n_H^{\alpha  i ;\beta j}
\mbox{index}(L_{i}^\alpha)
\mbox{index}(L_{j}^\beta)
&=&
\sum_{\hat i, \hat j}
\Omega_{\hat i\hat j}b_\alpha^{\hat i}b_\beta^{\hat j},
\label{anomalycondition5}
\eeqa
 where, following \cite{Sadov}, we have defined
\beqa
\mbox{tr}_{L_i^\alpha} F_\alpha^2&=&\mbox{index}L_i^\alpha~ 
\mbox{tr}_{\alpha}F^2_\alpha,\\
\mbox{tr}_{L_i^\alpha} F_\alpha^4
&=&
x_{L_i^\alpha}
\mbox{tr}_{\alpha}F^4_\alpha
+
y_{L_i^\alpha}
(\mbox{tr}_{\alpha}F^2_\alpha)^2
\eeqa
for some trace $\mbox{tr}_{\alpha}$ taken in a preferred representation
of $G_\alpha$. In the following we take the fundamental representation for 
this representation for $SU(N)$ or $SO(2N)$, 
{\bf 27} for $E_6$, ${\bf 56}$ for $E_7$ and {\bf 248}
 for $E_8$.

The anomaly (\ref{factorizedI8}) is  also canceled by this mechanism. 
Indeed, (\ref{factorizedI8}) is further written in a compact form: 
\beqa
{\hat I^{\rm total}}_{8}&=&\frac1{32}\left(
\frac12 K \mbox{tr}R^2 + D_u x^{(1)} + D_v x^{(2)}
\right)^2,
\eeqa
where $K$ is the canonical divisor of the Hirzebruch surface $\Fn$, and 
$D_u$, $D_v$  are the divisors of the sections $z=0,\infty$, 
respectively. The square on the right hand side 
is understood as an intersection product.
By choosing the divisor of the fiber $D_s$ 
and $D_v$ above 
as a basis,  $K$, $D_u$ and $D_v$ 
can be expressed in terms of component vectors: 
\beqa
K&=&-(2+n)D_s -2 D_v~\equiv~K^{\hat i}D_{\hat i},\n
D_u&=& n D_s+ D_v ~\equiv~D_u^{\hat i}D_{\hat i}, \n
D_v&\equiv&D_v^{\hat i}D_{\hat i}.
\eeqa
The intersection form is given by
\beqa
\Omega_{{\hat i}{\hat j}}&=&\left(
\begin{array}{cc}
D_s\!\!\cdot\!\! D_s& D_s\!\!\cdot\!\!  D_v\\
D_v\!\!\cdot\!\! D_s&D_v\!\!\cdot\!\!  D_v
\end{array}
\right)
~=~
\left(
\begin{array}{cc}
0& 1\\
1&-n
\end{array}
\right).
\eeqa
Then ${\hat I^{\rm total}}_{8}$ can be written in the form (\ref{I=OmegaXX})
with
\beqa
X^{\hat i}&=&\frac12 a^{\hat i} trR^2
+\sum_{m=1,2}\sum_{G_\alpha\in E_8^{(m)}}
2 b_\alpha^{(m)\hat i}\mbox{tr}_\alpha F_\alpha,\\
a^{\hat i}&=&K^{\hat i},\\
b^{(1)\hat i}_\alpha&=&\frac1{60 r_\alpha}D_u^{\hat i},\\
b^{(2)\hat i}_\alpha&=&\frac1{60 r_\alpha}D_v^{\hat i},
\eeqa
where
\beqa
r_\alpha&\equiv&\frac{\mbox{tr}_\alpha F_\alpha^2}{\mbox{Tr}_{E_8}F_\alpha^2}.
\eeqa
Thus F-theory on an elliptic Calabi-Yau over $\Fn$, which shares the same 
matter spectrum as that of $E_8\times E_8$ 
heterotic string on $K3$ with instanton numbers 
$(12+n,12-n)$, can be anomaly-free also by this anomaly cancellation mechanism.

\subsection{Multiple singularities and anomalies}
As we already mentioned, the F-theory compactification on an elliptic Calabi-Yau 
over $\Fn$, is dual to the $E_8\times E_8$ heterotic string compactification on $K3$ with 
instanton numbers $(12+n,12-n)$. More precisely, suppose that 
$G \times H$ is a direct product maximal subgroup of, say, the first factor of 
$E_8$, such that (1) $G$ is simple and simply-laced, and 
(2) $H$ is semi-simple.
We assume that $H$ has $12+n$ 
instantons so the unbroken gauge group from this $E_8$ 
is $G$. For these cases 
the massless spectra of heterotic string and the dual geometries of 
F-theory are summarized in appendix B. 
As it is shown there, for each such pair $(G,H)$, one can find a specialized 
Weierstrass form of the elliptic fiberation such that\\
\\
\noindent
(1) 
The total number of the deformation parameters of the curve that 
preserve the particular singularity structure is precisely equal to the 
number of neutral hypermultiplets (arising from this $E_8$) 
computed by the index theorem in heterotic string theory.\\
\\
\noindent
(2)  
The leading order term in $z$ of the discriminant of the Weierstrass form 
factorizes, and the degree in $z'$ of each factor again coincides with 
the number of charged hypermultiplets obtained by the index theorem. 
\\

Therefore, at least for this class of F-theory compactifications, there is 
no anomaly since they share the same massless matter contents as 
those of heterotic strings on $K3$. 
This is the situation where the both theories are at ``generic" points 
in the moduli spaces. On the other hand, suppose that the F-theory curve  
is deformed in such a way that more than one factor of the leading order 
term of the discriminant comes to share a common zero locus, at which 
the singularity is more enhanced than the ones occurring at the ordinary 
matter loci at generic points in the moduli space. 
As we discussed at the end of the previous section, such a multiple singularity 
supports more BPS string junctions than when the discriminant loci are 
split apart. Since for all the cases listed in TABLE \ref{charged} a half of 
the localized string junctions at the enhanced point precisely correspond to the 
matter representations predicted from the heterotic string analysis, one may 
also expect that the additional string junctions appearing at the multiple 
point will also play a role to the generation of massless matter in F-theory.   

However, the anomaly cancellation condition forbids such net increase of 
chiral matter. For the Green-Schwarz mechanism to work, the total anomaly 
polynomial must factorize into the form (\ref{factorizedI8}) or (\ref{I=OmegaXX}).
This requires the absence of the $Y_4$ term in ${\hat I^{\rm total}}_{8}$,
and that imposes the constraint:
\beqa
n_H - n_V &=& 273 - 29 n_T, 
\eeqa
as we saw previously.  Thus as long as the number of the tensor multiplets 
is one, as is so for the smooth heterotic compactifications, the only possible 
change in the number of the hypermultiplets is one associated with 
the simultaneous change in the number of the vector multiplets, that is, 
the Higgs mechanism. In the present case, however, there is no gauge symmetry 
enhancement to expect at the multiple singularity, so this does not happen.

Also, even if one allows to change the number of tensor multiplets, the anomalies 
from the net increase of the hypermultiplets cannot be canceled. This is because 
the coefficient of $n_T$ in (\ref{H-V=273-29T}) is {\em minus} 29, and hence 
the increase of $n_T$ means the {\em decrease} of the hypermultiplets.

The total number of chiral matter is also constrained by geometry. 
In the generalized Green-Schwarz mechanism, the change 
in the number of the tensor hypermultiplets means 
the change in the self-intersection number of the canonical class $K$ of the base 
manifold of the elliptic fiberation; see (\ref{anomalycondition1}).
In the present case, the base is a Hirzebruch 
surface. The canonical class can change if the surface is blown up at some points.
Suppose that the Hirzebruch surface is blown up at a point, the 
canonical class is changed to \cite{Matsuzawa}
\beqa
K&=&-(2+n)D_s -2 D_v\n
&\rightarrow & -(2+n)D_s -2 D_v + e_1,
\eeqa
where $e_1$ is the exceptional divisor that has arisen due to the blow up. 
Since its intersection pairing is 
\beqa
e_1\cdot e_1=-1,~~~D_s\cdot e_1=D_v\cdot e_1=0,
\eeqa
the self-intersection $K\cdot K$ {\em decreases} from eight to seven,
which also implies that there arises more tensor multiplets and 
less hypermultiplets are allowed to exist.\footnote{Such a transition was 
first considered in \cite{WittenMandF}.
Colliding singularities in   
F-theory on a blown-up Hirzebruch were studied in \cite{BJ}.}

Therefore, in any case, any net change of the total number of chiral matter 
is inconsistent with anomaly cancellation. Is there any transition of geometry 
without any change of the total number of hypermultiplets before and after 
the transition? In fact, an example of such a transition to special points in 
the moduli space has already been found in \cite{Tani},
where the branes have some multiple singularities
and at the same time the theory remains anomaly free. 
We will discuss this in the next section.

\section{Anomaly-free multiple singularities}
\subsection{Enhancement from $SU(5)$ to $SO(12)$}
The curve found in \cite{Tani} is one which has an $SU(5)=I_5$ 
singularity at $z=0$, and also parameterized by (\ref{genericSU(5)fandg}),
except that $q_{n+6}$ is further specialized to the form
\beqa
q_{n+6}&=&h_{n+2}q_4
\eeqa
for some fourth-order polynomial $q_4$ in $z'$.\footnote{Examples of 
multiple singularity enhancement from $SU(5)$ to $SO(12)$, $E_6$ or $E_7$ 
were more recently considered in \cite{MorrisonTaylor}.} 
This means that 
all the roots of the equation $h_{n+2}$ are also ones of $q_{n+6}$. 
In this particular case we have
\beqa
f(z,z')&=&-3 h_{n+2}^4
+12 z h_{n+2}^2
   H_{n+4}
+z^2 \left(12 q_4 h_{n+2}^2-12 H_{n+4}^2\right)
+z^3 f_{n+8}
+\cdots,
\n
g(z,z')&=&
2 h_{n+2}^6
-12 z h_{n+2}^4
   H_{n+4}
   +z^2
   h_{n+2}^2\left(24  H_{n+4}^2-12 q_4 h_{n+2}^2\right) 
   \n
&&+z^3 \left(-f_{n+8}
   h_{n+2}^2+24 q_4 h_{n+2}^2 H_{n+4}-16 H_{n+4}^3\right)
+z^4 \left(2 f_{n+8} H_{n+4}+12 q_4^2 h_{n+2}^2\right)\n
&&+z^5 g_{n+12}+\cdots,
\n
\Delta&=&
108 z^5 h_{n+2}^6 \left(-2 q_4 f_{n+8}+g_{n+12}-24 q_4^2 H_{n+4}\right)\n
&&-9 z^6 h_{n+2}^4 \left(-96 q_4 f_{n+8} H_{n+4}+f_{n+8}^2+72 g_{n+12} H_{n+4}+96
   q_4^3 h_{n+2}^2-1152 q_4^2 H_{n+4}^2\right)\n
&& +36 z^7 
h_{n+2}^2 (30 q_4^2 f_{n+8} h_{n+2}^2-24 q_4 f_{n+8}
   H_{n+4}^2+f_{n+8}^2 H_{n+4}-18 q_4 g_{n+12} h_{n+2}^2\n
   &&~~~~~~~~~~~~~~~~~+36 g_{n+12}
   H_{n+4}^2+432 q_4^3 h_{n+2}^2 H_{n+4}-288 q_4^2 H_{n+4}^3)  \n
&&-18 z^8 (3 f_{n+8} g_{n+12} h_{n+2}^2-72 q_4^2 f_{n+8} h_{n+2}^2 H_{n+4}-8
   q_4 f_{n+8}^2 h_{n+2}^2+2 f_{n+8}^2 H_{n+4}^2\n
   &&~~~~~~~~-72 q_4 g_{n+12} h_{n+2}^2
   H_{n+4}+48 g_{n+12} H_{n+4}^3-216 q_4^4 h_{n+2}^4)+\cdots.
\eeqa
We see that the coefficient of the leading order term of $\Delta$ has been 
changed to the form:
\beqa
\Delta&=&108 z^5 h_{n+2}^6 P_{n+12}+\cdots,
\n
P_{n+12}&\equiv&
-2 q_4 f_{n+8}+g_{n+12}-24 q_4^2 H_{n+4}
   \label{SU(5)toSO(12)Delta}
\eeqa
from $h_{n+2}^4 P_{3n+16}$ 
(\ref{genericSU(5)Delta}) 
for the generic $SU(5)$ curve.
If $h_{n+2}$ vanishes, then $f$ and $g$ start from $O(z^2)$ and $O(z^3)$,
respectively, and $\Delta$ vanishes all the way up to $O(z^7)$ with $O(z^8)$ 
being the first nonvanishing term.  This is a $D_6=I^*_2$ singularity, 
which means that the curve has $n+2$ points with multiple singularity
enhancement $SU(5)\rightarrow SO(12)$.

Note that this is another case of a collision of the loci of $h_{n+2}$ and $P_{3n+16}$
discussed in section II. To see this we set $h_{n+2}=0$ in 
$P_{3n+16}$ (\ref{genericSU(5)Delta}) to find that 
\beqa
P_{3n+16}&\sim &-24H_{n+4} q_{n+6}^2.
\label{Hp^2}
\eeqa 
Thus if either of $H_{n+4}$ or $q_{n+6}$ vanishes, there occurs 
a collision
\footnote{This collision is not the kind of one that needs a blowup 
on the base, unlike the cases discussed in \cite{BJ}. An extra tensor multiplet 
would make the theory anomalous in the present case as we saw 
at the end of the previous section.}. 
The former case was discussed in section II, where the singularity was 
enhanced to $E_6$%
\footnote{As we will see below, the fact that $q_{n+6}$ is being 
squared is important since it means a simultaneous degeneration of 
two loci of $P_{3n+16}$.}
.

In the present case, the BPS junctions are
the ones corresponding to the homogeneous K\"{a}hler manifold $SO(12)/(SU(5)\times U(1)^2)$:
\beqa
{\bf 10}(SO(10))\oplus{\bf 10}(SU(5))&=&{\bf 5}\oplus{\bf \bar 5}\oplus{\bf 10}
\eeqa
plus one ${\bf 1}$ from the extra Cartan subalgebra. 
Since ${\bf 5}$ and ${\bf \bar 5}$ 
are indistinguishable in six dimensions, 
we have
\beqa
{\bf 5}\oplus{\bf  5}\oplus{\bf 10}\oplus{\bf 1}
\eeqa
residing at each zero of $h_{n+2}$ ($SO(10)$ point).
Thus
the hypermultiplets at the brane intersections 
are 
\beqa
(n+2)\left({\bf 5}\oplus{\bf 5}\oplus{\bf 10}\oplus{\bf 1}\right) \oplus (n+12){\bf 5}
&=&(n+2){\bf 10}\oplus(3n+16){\bf 5}\oplus(n+2){\bf 1},
\eeqa 
where the $(n+12)$  {\bf 5}'s on the left hand side come from the zeros of 
$P_{n+12}$
(\ref{SU(5)toSO(12)Delta}).
In addition, there are singlets from the complex structure moduli; their number 
is determined by the degrees of freedom of the polynomials
\beqa
\mbox{$h_{n+2}$, $H_{n+4}$, $q_4$, $f_{n+8}$ and $g_{n+12}$},
\eeqa
which yield 
\beqa
(n+3)+(n+5)+5+(n+9)+(n+13)-1&=&4n+34
\label{sumofdegreesSO(12)}
\eeqa
more {\bf 1}'s, and hence $5n+36$ singlets in all.
Thus the matter spectrum coincides with (\ref{SU(5)curvespectrum})
and hence 
is unchanged from that for the generic unbroken 
$SU(5)$ curve we saw in section II, and therefore the theory remains anomaly-free!

\begin{figure}[h]%
\mbox{\hskip 0ex
\includegraphics[height=0.45\textheight]{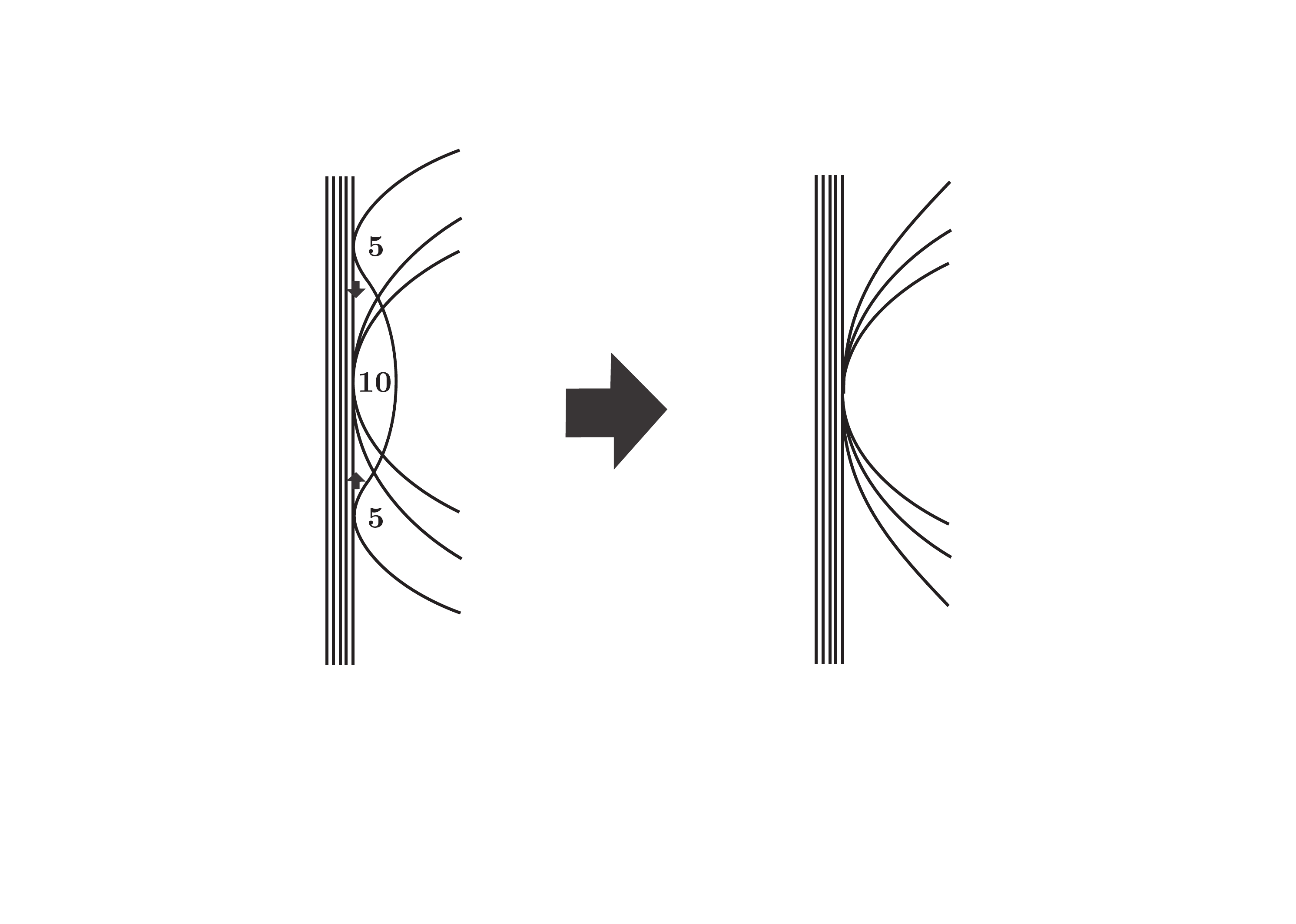}}\\
\vskip -15ex
\mbox{(a)$SU(5)\rightarrow SO(12)$}\\
\mbox{\hskip 0ex
\includegraphics[height=0.45\textheight]{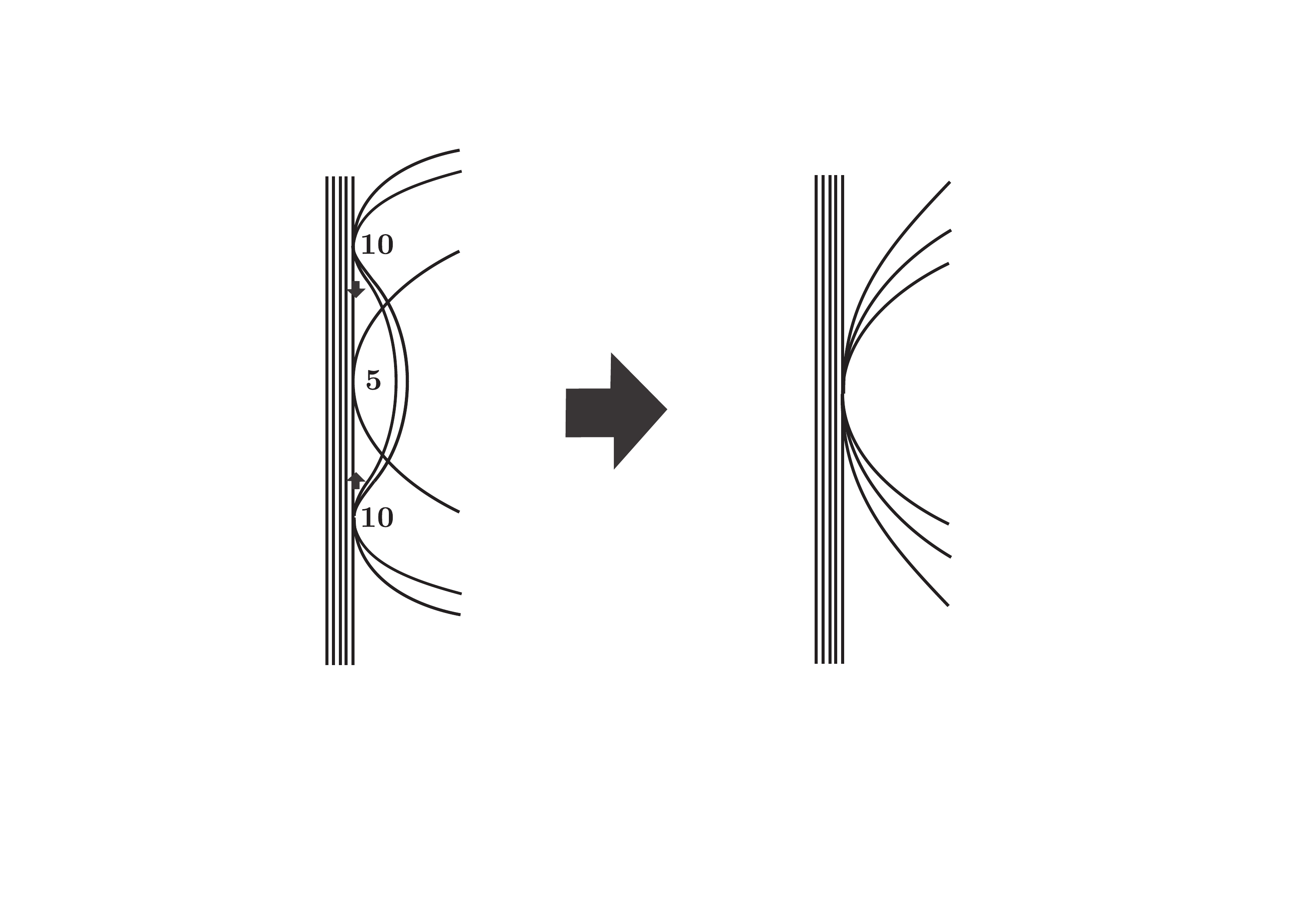}
}
\\
\vskip -15ex
\mbox{(b)$SU(5)\rightarrow E_6$}
\caption{\label{Anomalyfreemultiplesingularities} Anomaly-free multiple singularities. }
\end{figure}

How can this happen despite the extra ${\bf 5}$ at each zero locus of $h_{n+2}$? 
We can see this by noticing that the degree of the other factor of the leading 
term of the discriminant is changed to $n+12$ from $3n+16$ for the generic 
case. That is, $2n+4$ of $3n+16$ loci of ${\bf 5}$ have {\em pairwise degenerated} 
into $n+2$ pairs and simultaneously coalesced with the locus of ${\bf 10}$
(FIG.\ref{Anomalyfreemultiplesingularities}(a))!
Thus the total number of charged matter is unchanged. It is also remarkable 
that the balance of the neutral hypermultiplets is maintained before and after 
the multiple singularity enhancement; the emergence of the $n+2$ extra singlets %
at the singularity is precisely compensated by the decreased amount of complex 
structure moduli for the restricted geometry
\footnote{A similar anomaly cancellation can be seen for $SU(3)$ curves with multiple singularities. 
The generic $SU(3)$ curve (see TABLE \ref{neutral}) has the discriminant $\Delta = h_{n+2}^3 P_{6n+18} z^3 + \cdots$.
At each root of $P_{6n+18}$, 
the enhancement $I_3 \rightarrow I_4$  ($SU(3) \rightarrow SU(4)$) 
occurs and a ${\bf 3}$ appears, 
giving in all $(6n+18){\bf 3}$.
At a root of $h_{n+2}$, the fiber type changes as $I_3 \rightarrow IV$ ($SU(3) \rightarrow SU(3)$), where a {\bf B}-brane intersects the three {\bf A}-branes. At this point, no extra BPS string junction can
exist and hence no hypermultiplet appears. 
Specializing the generic curve in TABLE \ref{neutral} to 
$H_{2n+6}=h_{n+2}q_{n+4}$, we obtain multiple singularities \cite{Tani}.
The discriminant changes to $\Delta = h_{n+2}^6 P_{3n+12} z^3 +\cdots$.
It means that among the $6n+18$ roots of $P_{6n+18}$, $3n+6$ roots 
{\em triply degenerate} into $n+2$ sets and coalesce with zeros of $h_{n+2}$, 
yielding the multiple singularities.
At the remaining $3n+12$ roots of $P_{3n+12}$, $(3n+12){\bf 3}$ appear.
The decrease of $(3n+6){\bf 3}$
is precisely compensated by the ${\bf 3}$s at the multiple singularities of $h_{n+2}$.
In fact, at each root of $h_{n+2}$, enhancement $I_3 \rightarrow I_0^*$ 
($SU(3) \rightarrow SO(8)$) occurs and hypermultiplets in ${\bf 3}\oplus{\bf 3}\oplus{\bf 3}\oplus{\bf 1}$ appear.
Note that this last ${\bf 1}$ ($(n+2){\bf 1}$ in all) just 
compensates the decrease of 
$n+2$ neutral hypermultiplets due to 
the decrease of the complex structure moduli 
$H_{2n+6}\rightarrow q_{n+4}$
 via the specialization.}.

 The geometry considered in this section is the one with a
{\em maximal} number of multiple enhanced points from $SU(5)$ to $SO(12)$;
one may equally well consider the case where, for arbitrary integer $r$ 
$(0\leq r\leq n+2)$, 
$2r$ of $3n+16$ loci of ${\bf 5}$ 
pairwise merge with $r$ ${\bf 10}$ loci while the rest of ${\bf 5}$ loci remain as they are.
It is easy to see also in this case the numbers of 
charged and neutral matter do not change before and after the coalesce 
of the singularities.

Conversely, if the extra matter did {\em not} arise at the multiple singularity enhancement 
with the simultaneous degeneration of matter loci as above,  the balance of 
the matter multiplets  (\ref{H-V=273-29T}) would be lost and the theory 
would become anomalous. Thus the absence of anomalies requires here 
the generation of extra matter at this multiple singularity.

\subsection{Enhancement from $SU(5)$ to $E_6$}
Having understood how an anomaly-free multiple singularity enhancement 
can be realized, we can now find curves with other types of multiple singularity 
enhancement. Let us reexamine in this section the singularity enhancement 
$SU(5)\rightarrow E_6$ considered in section II.  As we saw there, this 
happens when $h_{n+2}$ and $H_{n+4}$ have a common zero locus. 

We first examine the case when $H_{n+4}$ takes the form 
\beqa
H_{n+4}&=&h_{n+2}H_2 \label{Hn+4single}
\eeqa
Although this particular form of $H_{n+4}$ indeed creates $n+2$ $E_6$ 
points, the extra hypermultiplets such as those that 
were described at the end of section II do {\em not} arise.
Indeed, with the form of $H_{n+4}$ (\ref{Hn+4single}), the discriminant takes 
the form 
\beqa
\Delta&=&h_{n+2}^5 P_{2n+14} z^5+\cdots,
\eeqa
but the $E_6$ multiple singularity with extra hypermultiplets 
is supposed to have two ${\bf 10}$'s at each locus so 
there are too many ${\bf 10}$'s to cancel anomalies.

Therefore, to keep the number of {\bf 10} unchanged, we instead set 
\beqa
h_{n+2}&=&h_{\frac{n+2}2}^2,\n
H_{n+4}&=&h_{\frac{n+2}2} H_{\frac{n+6}2}. \label{Hn+4double}
\eeqa
for some $h_{\frac{n+2}2}$, where $n+2$ is assumed to be divisible by two.
With (\ref{Hn+4double}) the discriminant reads
\beqa
\Delta&=&h_{\frac{n+2}2}^9 P_{\frac{5n+30}2} z^5+\cdots.
\eeqa
We see that the $n+2$ roots of the equation $h_{n+2}=0$ pairwise degenerate 
into $\frac{n+2}2$ double roots, each of which merges with a root of 
$P_{3n+16}$ (FIG.\ref{Anomalyfreemultiplesingularities}(b)).

The relevant homogeneous K\"{a}hler manifold in this case is 
$E_6/(SU(5)\times U(1)^2)$ whose $SU(5)$ representations are 
\beqa
{\bf 16}(SO(10))\oplus{\bf 10}(SU(5))&=&{\bf 10}\oplus{\bf \bar 5}\oplus{\bf 1}\oplus{\bf 10},
\eeqa
and there is another ${\bf 1}$ from the Cartan subalgebra. 
Thus the hypermultiplets coming from the brane intersections are 
\beqa
{\textstyle \frac{n+2}2} \left(2\cdot{\bf 10}\oplus{\bf 5}\oplus 2\cdot{\bf 1}\right) 
\oplus {\textstyle \frac{5n+30}2} {\bf 5}
&=&(n+2){\bf 10}\oplus(3n+16){\bf 5}\oplus (n+2){\bf 1}.
\label{multipleE6spectrum}
\eeqa 
On the other hand the decrease of the degrees of freedom of the polynomials 
is $\frac{n+2}2$ from $h_{n+2}\rightarrow h_{\frac{n+2}2}$ and 
$\frac{n+2}2$ from $H_{n+4}\rightarrow H_{\frac{n+6}2}$, in total $n+2$ again. 
This compensates the extra $n+2$ singlets in (\ref{multipleE6spectrum}), 
and the theory after this transition also remains to be anomaly-free.

Although we have for simplicity considered the case with maximally possible 
multiple $E_6$ points with $n+2$ even, we may similarly consider the case with 
less $E_6$ points and/or $n+2$ odd. All that we require is that a pair of degenerating 
loci of $h_{n+2}$ and a single locus of $P_{3n+16}$ merge together 
per one $E_6$ multiple singularity. In this way the number of charged hypermultiplets 
is conserved, but again, the fact that the number of new singlets also matches the 
decrease of the complex structure moduli is rather nontrivial.

\subsection{Enhancement from $SU(5)$ to $E_7$}
Let us now consider a singularity enhancement in which the rank of 
the Lie algebra characterizing the singularity jumps up by more than 
two. The enhancement from $SU(5)$ to $E_7$ is of particular interest 
because it is relevant to the F-theory realization \cite{FFamilyUnification} 
of the Kugo-Yanagida $E_7/(SU(5)\times U(1)^3)$ family unification model
\cite{KY}.

The Kodaira classification tells us that the $E_7=III^*$ singularity occurs 
when ${\rm ord}f=3$, ${\rm ord}g\geq 5$ and ${\rm ord}\Delta=9$.
We can see from appendix A that this happens when 
$h_{n+2}$, $H_{n+4}$ and $q_{n+6}$ all simultaneously vanish.
The homogeneous K\"ahler manifold for this singularity is 
$E_7/(SU(5)\times U(1)^3)$ with the following $SU(5)$ representations:
\beqa
{\bf 27}(E_6)\oplus{\bf 16}(SO(10))\oplus{\bf 10}(SU(5))&=&
({\bf 16}(SO(10)\oplus{\bf 10}(SO(10))\oplus{\bf 1})\n
&&\oplus{\bf 16}(SO(10))\oplus{\bf 10}(SU(5))\n
&=&
3\cdot{\bf 10}\oplus 4\cdot{\bf 5}\oplus 3\cdot{\bf 1},
\eeqa
where in the last line we have made no distinction between ${\bf 5}$
and ${\bf\bar 5}$. In addition, we have, this time, two ${\bf 1}$'s from 
the Cartan subalgebra. In all, three {\bf 10}'s, four {\bf 5}'s and 
five singlets are supposed to arise at each multiple $E_7$ singularity. 
Thus, in order for the anomalies to cancel, we need to have  
three loci of $h_{n+2}$ and four loci of $P_{3n+16}$ to 
simultaneously degenerate and join together, per one $E_7$ singularity.
This is achieved, again for the maximal case, at the special points in the 
moduli space as follows:
\beqa
h_{n+2}&=&h_{\frac{n+2}3}^3,\n
H_{n+4}&=&h_{\frac{n+2}3}^2 H_{\frac{n+8}3}, \n
q_{n+6}&=&h_{\frac{n+2}3} q_{\frac{2n+16}3}. \label{E7rescalings}
\eeqa
for some $h_{\frac{n+2}3}$, $H_{\frac{n+8}3}$ and $q_{\frac{2n+16}3}$, 
where $n+2$ is assumed to be divisible by 
three in this case. The non-maximal case and/or the case in which $n+2$ 
is not 0 mod 3 are treated similarly. With (\ref{E7rescalings}) 
the discriminant becomes 
\beqa
\Delta&=&h_{\frac{n+2}3}^{16} P_{\frac{5n+40}3} z^5+\cdots.
\eeqa
The total number of ${\bf 5}$ is thus 
\beqa
4\times \frac{n+2}3+\frac{5n+40}3&=&3n+16,
\eeqa
which is a correct value. Also the decrease of the degrees of freedom 
of the polynomials is 
\beqa
2\times \frac{n+2}3+2\times \frac{n+2}3+1\times \frac{n+2}3&=&5\times \frac{n+2}3,
\eeqa
which match the five singlets residing at each of the $\frac{n+2}3$ $E_7$ points.

%

\subsection{Enhancement from $SU(5)$ to $E_8$}
The final example of anomaly-free singularity enhancement we consider in 
this paper is the one from $SU(5)$ to $E_8$.
This type of multiple singularity may also be used for particle physics 
model building because the $D=4$ supersymmetric nonlinear sigma model 
with $E_8/(SU(5)\times U(1)^4)$ as the target also yields net three 
chiral generations. Furthermore, it was pointed out \cite{lopsided} that 
this coset may also give rise to three sets of nonchiral singlet pairs 
needed in a scenario proposed by Sato and Yanagida  \cite{SatoYanagida}
explaining the Yukawa hierarchies and large lepton-flavor mixings 
by the Frogatt-Nielsen mechanism
\footnote{In fact, the $E_8$ 
curve given in the original version of  \cite{lopsided} did not take 
account of the simultaneous degenerations of loci and hence was 
anomalous as a six-dimensional theory. 
A revised version is in preparation.}.

The spectrum of $E_8/(SU(5)\times U(1)^4)$ is
\beqa
5\cdot{\bf 10}\oplus 10\cdot{\bf 5}\oplus 10\cdot{\bf 1}.
\eeqa
With the additional three singlets from the Cartan subalgebra, 
in all 
\beqa
5\cdot{\bf 10}\oplus 10\cdot{\bf 5}\oplus 13\cdot{\bf 1} 
\eeqa
reside at each $E_8$ point\footnote{At first sight it seems that 
the $D=4$, ${\cal N}=1$ supersymmetric nonlinear sigma model 
with this target space may have 
five generations. However, one can show that \cite{IKK} 
it is not possible to choose 
a so-called ``Y-charge'', a $U(1)$ charge that determines the complex 
structure of the coset space, in such a way that all the five ``flavors"
may have the same chirality.  See \cite{IKK} for more detail.}.

We can also find an anomaly-free curve with these $E_8$ multiple 
singularities.
We again only present the case where $n+2$ is divisible by five and 
all the $h_{n+2}$ loci turn into the $E_8$ singularities: 
\beqa
h_{n+2}&=&h_{\frac{n+2}5}^5,\n
H_{n+4}&=&h_{\frac{n+2}5}^4 H_{\frac{n+12}5}, \n
q_{n+6}&=&h_{\frac{n+2}5}^3 q_{\frac{2n+24}5},\n
f_{n+8}&=&h_{\frac{n+2}5}^2 f_{\frac{3n+36}5}
\label{E8rescalings}
\eeqa
for some $h_{\frac{n+2}5}$, $H_{\frac{n+12}5}$, $q_{\frac{2n+24}5}$ 
and $f_{\frac{3n+36}5}$.
%
%
Then  the discriminant reads 
\beqa
\Delta&=&h_{\frac{n+2}5}^{30} P_{n+12} z^5+\cdots.
\eeqa
We can similarly verify that the numbers of both charged and neutral hypermultiplets 
are the same as those at generic points in the moduli space.  
Therefore, also in this case, the theory is anomaly-free.

It is interesting to notice that the powers of $h_{\frac{n+2}5}$ factors in 
(\ref{E8rescalings}) are precisely the exponents of $SU(5)$, that is, 
the powers of the canonical class projectivized in the weighted projective 
bundle \cite{FMW},
of which (\ref{h's_and_a's}) are sections.
Perhaps this coincidence may be interpreted in terms of spectral 
covers of the dual heterotic string theory.

\section{Conclusions}
We have shown that multiple singularity enhancement 
can really occur in F-theory without causing imbalance of anomalies. 
We have considered concrete examples in F-theory compactifications 
on an elliptically fibered Calabi-Yau over a Hirzebruch surface $\Fn$.
Anomaly cancellation requires that there should be no net change 
in numbers of hypermultiplets after the coalesce of matter loci. 
We have presented such particular points in the moduli space in 
the case of unbroken $SU(5)$ gauge group, where the singularity 
is multiply enhanced to $SO(12)$, $E_6$, $E_7$ or $E_8$.

Although we have mainly considered the six-dimensional F-theory 
with $G=SU(5)$, it is natural to expect a similar anomaly-free 
transition to a configuration with multiple singularities to occur 
in four-dimensional compactifications and/or with 
other gauge groups. The original motivation to consider 
multiple singularity enhancement in F-theory was to construct 
``family unification'' particle physics models in string theory. 
But if the number of chiral matter does not change after the 
coalesce of singularities,  what is the use of the multiple 
singularities in string phenomenology model building?

To consider the multiple singularity enhancement in F-theory 
has at least three virtues:\\
\noindent
(1) In general, a special point in the moduli space can be an end point
of whatever flow in the moduli space after the supersymmetry is 
broken and potentials are generated; if it is not a special 
point, there is no reason for the flow to stop at that point. \\
\noindent
(2) The multiple singularity may occur, in principle, in any elliptic 
Calabi-Yau manifold.  Since the structure is universal, it may offer a 
potential ubiquitous mechanism for generating three generations 
of flavors in the framework of F-theory. \\
\noindent
(3) Last but not least, the homogeneous K\"{a}hler structure of 
the spectrum of the multiple singularity is naturally endowed with 
conserved $U(1)$ charges. This may also be useful for particle physics 
model building.

It would be extremely interesting to extend the analysis done in 
this paper to four dimensions.

\section*{Acknowledgments} 

We thank K.~Mohri for useful discussions. The work of SM is supported by Grant-in-Aid 
for Scientific Research (C) \#25400285 and (A) \#26247042 from The
Ministry of Education, Culture, Sports, Science and Technology of Japan.

\section*{Appendix A $SU(5)$ curve}
\label{appendixA} 
\noindent
\beqa
f(z,z')&=&-3 h_{n+2}^4+12 z h_{n+2}^2 H_{n+4}+z^2 \left(12 h_{n+2} q_{n+6}-12
   H_{n+4}^2\right)+z^3 f_{n+8}+f_8 z^4+z^5 f_{8-n}\n
   &&+O\left(z^{6}\right),\\
g(z,z')&=&2 h_{n+2}^6-12 z~ h_{n+2}^4 H_{n+4}
%
\n
&&+12 z^2 h_{n+2}^2(2 
   H_{n+4}^2- h_{n+2} q_{n+6})\n
   &&+z^3 \left(-f_{n+8} h_{n+2}^2+24 h_{n+2}
   H_{n+4} q_{n+6}-16 H_{n+4}^3\right)\n
   &&+z^4 \left(-f_8 h_{n+2}^2+2 f_{n+8}
   H_{n+4}+12 q_{n+6}^2\right)+z^5 g_{n+12}+g_{12} z^6+O\left(z^{7}\right),\\
\Delta&=&
108 z^5 h_{n+2}^4 (-2 f_8 h_{n+2}^2 H_{n+4}-2 f_{n+8} h_{n+2}
   q_{n+6}+f_{8-n} h_{n+2}^4+g_{n+12} h_{n+2}^2\n
&&~~~-24 H_{n+4} q_{n+6}^2)\n
&&-9 z^6 h_{n+2}^2 (-96 f_{n+8} h_{n+2} H_{n+4} q_{n+6}+96 f_{8-n}
   h_{n+2}^4 H_{n+4}-144 f_8 h_{n+2}^2 H_{n+4}^2\n
&&~~~+24 f_8 h_{n+2}^3
   q_{n+6}+f_{n+8}^2 h_{n+2}^2+72 g_{n+12} h_{n+2}^2 H_{n+4}-12 g_{12}
   h_{n+2}^4+96 h_{n+2} q_{n+6}^3\n
&&~~~-1152 H_{n+4}^2 q_{n+6}^2)\n
&&-18 z^7 (-120 f_8 h_{n+2}^3 H_{n+4} q_{n+6}+48 f_{n+8} h_{n+2} H_{n+4}^2
   q_{n+6}-144 f_{8-n} h_{n+2}^4 H_{n+4}^2\n
&&~~~+144 f_8 h_{n+2}^2 H_{n+4}^3-2
   f_{n+8}^2 h_{n+2}^2 H_{n+4}+48 f_{8-n} h_{n+2}^5 q_{n+6}-60 f_{n+8}
   h_{n+2}^2 q_{n+6}^2\n
&&~~~+f_8 f_{n+8} h_{n+2}^4+36 g_{12} h_{n+2}^4 H_{n+4}-72
   g_{n+12} h_{n+2}^2 H_{n+4}^2+36 g_{n+12} h_{n+2}^3 q_{n+6}\n
&&~~~-864 h_{n+2}
   H_{n+4} q_{n+6}^3+576 H_{n+4}^3 q_{n+6}^2)\n
&&+9 z^8 (-6 f_{n+8} g_{n+12} h_{n+2}^2+384 f_{8-n} h_{n+2}^3 H_{n+4}
   q_{n+6}-384 f_8 h_{n+2} H_{n+4}^2 q_{n+6}\n
&&~~~-384 f_{8-n} h_{n+2}^2
   H_{n+4}^3+20 f_8 f_{n+8} h_{n+2}^2 H_{n+4}+120 f_8 h_{n+2}^2 q_{n+6}^2\n
&&~~~+16
   f_{n+8}^2 h_{n+2} q_{n+6}-f_8^2 h_{n+2}^4-8 f_{8-n} f_{n+8} h_{n+2}^4+144
   f_{n+8} H_{n+4} q_{n+6}^2\n
&&~~~+192 f_8 H_{n+4}^4-4 f_{n+8}^2 H_{n+4}^2+144
   g_{n+12} h_{n+2} H_{n+4} q_{n+6}+144 g_{12} h_{n+2}^2 H_{n+4}^2\n
&&~~~-72 g_{12}
   h_{n+2}^3 q_{n+6}-96 g_{n+12} H_{n+4}^3+432 q_{n+6}^4)\n
&&+2 z^9 (-27 g_{12} f_{n+8} h_{n+2}^2-27 f_8 g_{n+12} h_{n+2}^2+54 f_{n+8}
   g_{n+12} H_{n+4}\n
&&~~~-1728 f_{8-n} h_{n+2} H_{n+4}^2 q_{n+6}+72 f_8^2 h_{n+2}^2
   H_{n+4}+144 f_{8-n} f_{n+8} h_{n+2}^2 H_{n+4}\n
&&~~~+864 f_{8-n} h_{n+2}^2
   q_{n+6}^2+144 f_8 f_{n+8} h_{n+2} q_{n+6}-36 f_8 f_{8-n} h_{n+2}^4+864
   f_{8-n} H_{n+4}^4\n
&&~~~-144 f_8 f_{n+8} H_{n+4}^2+2 f_{n+8}^3+648 g_{12} h_{n+2}
   H_{n+4} q_{n+6}-432 g_{12} H_{n+4}^3\n
&&~~~+324 g_{n+12} q_{n+6}^2)\n
&&+3 z^{10} (-18 f_8 g_{12} h_{n+2}^2+36 g_{12} f_{n+8} H_{n+4}+96
   f_8 f_{8-n} h_{n+2}^2 H_{n+4}+48 f_8^2 h_{n+2} q_{n+6}\n
&&~~~+96 f_{8-n} f_{n+8}
   h_{n+2} q_{n+6}-12 f_{8-n}^2 h_{n+2}^4-48 f_8^2 H_{n+4}^2-96 f_{8-n}
   f_{n+8} H_{n+4}^2\n
&&~~~+4 f_8 f_{n+8}^2+216 g_{12} q_{n+6}^2+9
   g_{n+12}^2)+O(z^{11}).
\eeqa

\section*{Appendix B}
\label{appendixB} 
\noindent
In this appendix we summarize the details of  the correspondence between 
massless matter spectra of $E_8\times E_8$ heterotic string theory 
compactified on $K3$ and geometric data of elliptically fibered Calabi-Yau 
three-fold over Hirzebruch surfaces on which F-theory is compactified.

Let $E_8^{(1)}$ ($E_8^{(2)}$) be the first (second) factor of $E_8\times E_8$ 
and $G^{(m)}\times H^{(m)}$ $(m=1,2)$ be a direct product maximal subgroup 
of $E_8^{(m)}$ $(m=1,2)$. We assume that $H^{(1)}$ ($H^{(2)}$) has $12+n$ 
($12-n$) instantons. We restrict ourselves to the cases where 
(1) $G^{(m)}$ is simple and simply-laced, and (2) $H^{(m)}$ is semi-simple.
The massless spectrum of heterotic string can be computed \cite{GSWest} 
by the index theorem (\ref{indexTH}).

The TABLE \ref{neutral} shows the neutral matter spectrum of the heterotic 
string arising from $E_8^{(1)}$, the corresponding Weierstrass form of the 
F-theory curve and the independent polynomials which parametrize 
the curve. The subscripts denote the degrees of the polynomials 
in $z'$. For each pair of $G=G^{(1)}$ and $H=H^{(1)}$, the sum of the numbers 
of the coefficients of the independent polynomials, minus one which takes 
account of the overall rescaling, always coincides with the number of heterotic 
singlets obtained by the index theorem, as was verified in (\ref{sumofdegrees}) 
in section II. A similar result holds for the neutral 
matter from $E_8^{(2)}$ and the coefficients of the Weierstrass form 
$\sum_{i=5}^8 z^i f_{8+(4-i)n}(z')$ and 
$\sum_{i=7}^{12} z^i g_{12+(6-i)n}(z')$ which determine the singularity at $z=\infty$.

The TABLE \ref{charged} shows the spectrum of the charged hypermultiplets.
For each $(G,H)$, the leading order term in $z$ of the discriminant of the Weierstrass form 
factorizes, and the degree in $z'$ of each factor coincides with the number of 
charged hypermultiplets obtained by the index theorem. What representation 
occurs is related to the pattern of the singularity enhancement as explained 
in the text. We have also shown in the last column the corresponding divisor 
whose intersection number with $D_u$ (the divisor for the $z=0$ section), 
determined by the anomaly cancellation conditions 
(\ref{anomalycondition1})-(\ref{anomalycondition5}), 
gives the number of the charged hypermultiplets.

\begin{table}[bh]
\caption{Heterotic/F-theory duality: Neutral hypermultiplets. \label{neutral}}
\centering
\begin{tabular}{|c|c||c||c|c|c|}
\hline
$G$& $H$ 
&$\begin{array}{c}\mbox{heterotic/}\\
\mbox{neutral}\\ \mbox{matter}\end{array}$
&F-theory curve &indep't polynomials\\
\hline
$E_7$
&$SU(2)$
&$(2n+21){\bf 1}$
& $\begin{array}{c}f_{8+4n}=f_{8+3n}=f_{8+2n}=g_{12+6n}=g_{12+5n}\\
=g_{12+4n}=g_{12+3n}=g_{12+2n}=0\end{array}$
&$g_{n+12},f_{n+8}$
\\
\hline
$E_6$
&$SU(3)$
&$(3n+28){\bf 1}$
&$\begin{array}{c}f_{8+4n}=f_{8+3n}=f_{8+2n}=g_{12+6n}=g_{12+5n}\\
=g_{12+4n}=g_{12+3n}=0,~~g_{12+2n}=q_{n+6}^2\end{array}$
&$g_{n+12},f_{n+8},q_{n+6}$
\\
\hline
$SO(12)$
&$SO(4)$
&$(2n+18){\bf 1}$
&$\begin{array}{c}
f_{8+4n}\!=\!f_{8+3n}\!=\!g_{12+6n}\!=\!g_{12+5n}\!=\!g_{12+4n}\!=\! 0,\\
f_{8+2n}\!=\!-3H_{n+4}^2,~g_{12+3n}\!=\!2H_{n+4}^3,\\g_{12+2n}\!=\!-f_{n+8}H_{n+4},\\
~12(g_{12+n}+f_8H_{n+4})H_{n+4}+f_{n+8}^2=q_{n+8}^2\end{array}$
&$\begin{array}{c}q_{n+8},f_{n+8},H_{n+4}\\ \mbox{with}\\ \mbox{$n+4$ constraints}
\end{array}$
\\
\hline
$SO(10)$
&$SU(4)$
&$(4n+33){\bf 1}$
&$\begin{array}{c}
f_{8+4n}\!=\!f_{8+3n}\!=\!g_{12+6n}\!=\!g_{12+5n}\!=\!g_{12+4n}\!=\! 0,\\
f_{8+2n}\!=\!-3H_{n+4}^2,~g_{12+3n}\!=\!2H_{n+4}^3,\\g_{12+2n}\!=\!-f_{n+8}H_{n+4}+q_{n+6}^2\end{array}$
&$\begin{array}{c}g_{n+12},f_{n+8},q_{n+6},\\H_{n+4}  
\end{array}$
\\
\hline
$SO(8)$
&$SO(8)$
&$(6n+44){\bf 1}$
&$\begin{array}{c}
f_{8+4n}\!=\!f_{8+3n}\!=\!g_{12+6n}\!=\!g_{12+5n}\!=\!g_{12+4n}\!=\! 0,\\
4f_{8+2n}^3+27g_{12+3n}^2\!=\!j_{n+4}^2 k_{n+4}^2(j_{n+4}+k_{n+4})^2\end{array}$
&$\begin{array}{c}g_{2n+12},g_{n+12},f_{n+8},\\j_{n+4},k_{n+4}   
\end{array}$
\\
\hline
$SU(6)$
&$\begin{array}{c}SU(3)\\
\times SU(2)\end{array}$
&$(3n-r+21){\bf 1}$
&$\begin{array}{l}
\mbox{the same as $SU(5)$ with}\\
h_{n+2}\!=\! t_r \tilde h_{n+2-r},~
q_{n+6}\!=\! u_{r+4}\tilde h_{n+2-r},\\
H_{n+4}\!=\! t_r q_{n-r+4},\\
f_{n+8}\!=\! -12 u_{r+4}q_{n-r+4}+ t_r p_{n-r+8},\\
g_{n+12}\!=\!2u_{r+4}p_{n-r+8}-f_{8-n}h_{n+2}^2+f_8 H_{n+4}
\end{array}$
&$\begin{array}{c}t_r, \tilde h_{n+2-r}, u_{r+4},\\q_{n-r+4},p_{n-r+8}
\end{array}$
\\
\hline
$SU(5)$
&$SU(5)$
&$(5n+36){\bf 1}$
&$\begin{array}{l}
f_{8+4n}\!=\!-3 h_{n+2}^4,
g_{12+6n}\!=\!2 h_{n+2}^6,\\
g_{12+5n}\!=\! -12 h_{n+2}^4 H_{n+4},\\
f_{8+3n}\!=\!12 h_{n+2}^2 H_{n+4},\\
g_{12+4n}\!=\! h_{n+2}^2(12 H_{n+4}^2-f_{8+2n}),\\
g_{12+3n}\!=\! 2f_{8+2n}H_{n+4}+8H_{n+4}^3-f_{8+n}h_{n+2}^2\\
f_{8+2n}\!=\! -12H_{n+4}^2+12h_{n+2}q_{n+6}\\
g_{12+2n}\!=\! 12 q_{n+6}^2 + 2f_{8+n}H_{n+4} - f_8 h_{n+2}^2
\end{array}$
&$\begin{array}{c}h_{n+2},H_{n+4},\\
q_{n+6},f_{8+n},g_{12+n}  
\end{array}$
\\
\hline
\end{tabular}
\end{table}

\begin{table}[h]
\hskip -85ex(Cont'd)\\
\noindent
%
\centering
\begin{tabular}{|c|c||c||c|c|c|}
\hline
%
%
%
%
%
$SU(4)$
&$SO(10)$
&$\begin{array}{c}\mbox{\phantom{heterotic/}}\\
(8n+51){\bf 1}\\
\mbox{\phantom{singlets}}
\end{array}$
&$\begin{array}{l}
f_{8+4n}\!=\!-3 h_{n+2}^4,
g_{12+6n}\!=\!2 h_{n+2}^6,\\
g_{12+5n}\!=\! -12 h_{n+2}^4 H_{n+4},\\
f_{8+3n}\!=\!12 h_{n+2}^2 H_{n+4},\\
g_{12+4n}\!=\! h_{n+2}^2(12 H_{n+4}^2-f_{8+2n}),\\
g_{12+3n}\!=\! 2f_{8+2n}H_{n+4}+8H_{n+4}^3-f_{8+n}h_{n+2}^2
\end{array}$
&$\begin{array}{c}h_{n+2},H_{n+4},f_{8+2n},\\
f_{8+n},g_{12+2n},g_{12+n}  
\end{array}$
\\
\hline
$SU(3)$
&$E_6$
&$(12n+66){\bf 1}$
&$\begin{array}{l}
f_{8+4n}\!=\!-3 h_{n+2}^4,
g_{12+6n}\!=\!2 h_{n+2}^6,\\
g_{12+5n}\!=\! -12 h_{n+2}^3 H_{2n+6},\\
g_{12+4n}\!=\! 12 H_{2n+6}^2-h_{n+2}^2 f_{8+2n},\\
f_{8+3n}\!=\!12 h_{n+2} H_{2n+6},
\end{array}$
&$\begin{array}{c}h_{n+2},H_{2n+6},f_{8+2n},\\
f_{8+n},g_{12+3n},\\g_{12+2n},g_{12+n}  
\end{array}$
\\
\hline
$SU(2)$
&$E_7$
&$(18n+83){\bf 1}$
&$\begin{array}{l}
f_{8+4n}\!=\!-3 h_{2n+4}^2,
g_{12+6n}\!=\!2 h_{2n+4}^3,\\
g_{12+5n}\!=\! -h_{2n+4} f_{8+3n}
\end{array}$
&$\begin{array}{c}h_{2n+4},f_{8+3n},f_{8+2n},\\
f_{8+n},g_{12+4n},g_{12+3n},\\g_{12+2n},g_{12+n}  
\end{array}$
\\
\hline
\end{tabular}
\end{table}

\begin{table}[h]
\caption{Heterotic/F-theory duality: Charged hypermultiplets. \label{charged}}
\centering
\begin{tabular}{|c|c||c||c|c|c|}
\hline
$G$& $H$ & $\begin{array}{c}\mbox{heterotic/}\\
\mbox{charged}\\ \mbox{matter}\end{array}$ &   matter locus 
& $\begin{array}{c}\mbox{singularity}\\\mbox{enhancement}\end{array}$& divisor \\
\hline
$E_7$
&$SU(2)$
& $\frac{n+8}2 {\bf 56}$
&$f_{8+n}$
&$E_7\rightarrow E_8$
&$ -2K-\frac32 D_u$
\\
\hline
$E_6$
&$SU(3)$
& $(n+6){\bf 27}$
&$q_{n+6}$
&$E_6\rightarrow E_7$
& $ -3K-2 D_u$
\\
\hline
$SO(12)$
&$SO(4)$
& $\begin{array}{c}\frac{n+4}2{\bf 32}\\(n+8){\bf 12}\end{array}$
&$\begin{array}{c}H_{n+4}\\q_{n+8}\end{array}$
&$\begin{array}{c}SO(12)\rightarrow E_7\\ SO(12)\rightarrow SO(14) \end{array}$
&$\begin{array}{c}-K-\frac12 D_u\\-4K-3 D_u\end{array}$
\\
\hline
$SO(10)$
&$SU(4)$
& $\begin{array}{c}(n+4){\bf 16}\\(n+6){\bf 10}\end{array}$
&$\begin{array}{c}H_{n+4}\\q_{n+6}\end{array}$
&$\begin{array}{c}SO(10)\rightarrow E_6\\ SO(10)\rightarrow SO(12) \end{array}$
& $\begin{array}{c}-2K- D_u\\ -3K-2 D_u\end{array}$
\\
\hline
$SO(8)$
&$SO(8)$
& $\begin{array}{c}(n+4){\bf 8}_v\\(n+4){\bf 8}_s\\(n+4){\bf 8}_c\end{array}$
&$\begin{array}{c}j_{n+4}\\k_{n+4}\\j_{n+4}+k_{n+4}\end{array}$
&$\begin{array}{c}SO(8)\rightarrow SO(10)\\SO(8)\rightarrow SO(10)\\SO(8)\rightarrow SO(10) \end{array}$
& $\begin{array}{c}-2K- D_u\\-2K- D_u\\-2K- D_u\end{array}$
\\
\hline
$SU(6)$
&$\begin{array}{c}SU(3)\\
\times SU(2)\end{array}$
& $\begin{array}{c}\frac r2{\bf 20}\\(n-r+2){\bf 15}\\ 
(2n+r+16){\bf 6}\end{array}$
&$\begin{array}{c}t_r\\ \tilde h_{n+2-r}\\P_{2n+r+16}\end{array}$
&$\begin{array}{c}SU(6)\rightarrow E_6\\ SU(6)\rightarrow SO(12)\\SU(6)\rightarrow SU(7) \end{array}$
& $\begin{array}{c}-\frac r4 K -\frac r4 D_u\\(-1+\frac r2)K+\frac r2 D_u
\\-(8+\frac r2)K-(6+\frac r2)D_u\end{array}$
\\
\hline
$SU(5)$
&$SU(5)$
& $\begin{array}{c}(n+2){\bf 10}\\(3n+16){\bf 5}\end{array}$
&$\begin{array}{c}h_{n+2}\\P_{3n+16}\end{array}$
&$\begin{array}{c}SU(5)\rightarrow SO(10)\\ SU(5)\rightarrow SU(6) \end{array}$
& $\begin{array}{c}-K\\ -8K-5 D_u\end{array}$
\\
\hline
$SU(4)$
&$SO(10)$
& $\begin{array}{c}(n+2){\bf 6}\\(4n+16){\bf 4}\end{array}$
&$\begin{array}{c}h_{n+2}\\P_{4n+16}\end{array}$
&$\begin{array}{c}SU(4)\rightarrow SO(8)\\ SU(4)\rightarrow SU(5) \end{array}$
& $\begin{array}{c}-K\\ -8K-4 D_u\end{array}$
\\
\hline
$SU(3)$
&$E_6$
& $(6n+18){\bf 3}$
&$P_{6n+18}$
&$SU(3)\rightarrow SU(4)$
& $ -9K-3 D_u$
\\
\hline
$SU(2)$
&$E_7$
& $(6n+16){\bf 2}$
&$P_{6n+16}$
&$SU(2)\rightarrow SU(3)$
& $ -8K-2 D_u$
\\
\hline
\end{tabular}
\end{table}


\end{document}